# Analytical modeling of micelle growth. 1. Chain-conformation free energy of binary mixed spherical, wormlike and lamellar micelles

Krassimir D. Danov [a], Peter A. Kralchevsky [a,*], Simeon D. Stoyanov [b,c,d], Joanne L. Cook [e], Ian P. Stott [e]

[a] *Department of Chemical and Pharmaceutical Engineering, Faculty of Chemistry and Pharmacy, Sofia University, Sofia 1164, Bulgaria*

[b] *Unilever Research & Development Vlaardingen, 3133AT Vlaardingen, The Netherlands*

[c] *Laboratory of Physical Chemistry and Colloid Science, Wageningen University, 6703 HB Wageningen, The Netherlands*

[d] *Department of Mechanical Engineering, University College London, WC1E 7JE, UK*

[e] *Unilever Research & Development Port Sunlight, Bebington CH63 3JW, UK*

ABSTRACT

*Hypotheses:*
A quantitative molecular-thermodynamic theory of the growth of giant wormlike micelles of nonionic surfactants can be developed on the basis of a generalized model, which includes the classical 'phase separation' and 'mass action' models as special cases. The generalized model describes spherocylindrical micelles, which are simultaneously *multicomponent* and *polydisperse* in size.

*Theory:*
By analytical minimization of the free-energy functional we derived explicit expressions for the chain-extension and chain-end distribution functions in the hydrocarbon core of mixed micelles from two surfactants of different chainlengths.

*Findings:*
The hydrocarbon core of a two-component micelle is divided in two regions, *outer* and *inner*, where the ends of the shorter and longer chains are located. The derived analytical expression for the chain-conformation free energy implies that the mixing of surfactants with different chainlengths is always *nonideal* and *synergistic*, i.e. it leads to decrease of the micellar free energy and to enhancement of micellization and micelle growth. The derived expressions are applicable to surfactants with different headgroups (nonionic, ionic, zwitterionic) and to micelles of different shapes (spherical, wormlike, lamellar). The results can be incorporated in a quantitative theory of the growth of giant mixed micelles in formulations with practical applications in detergency.

* Corresponding author. Tel.: +359 2 962 5310
  E-mail address: pk@lcpe.uni-sofia.bg (P.A. Kralchevsky)

# 1. Introduction

Growth of wormlike micelles and other giant self-assembled aggregates has been observed in solutions containing single surfactant: nonionic [1,2,3]; zwitterionic [4,5], and ionic in the presence of added salt [6,7,8,9]. However, the growth of large micellar aggregates is most frequently observed in *mixed* surfactant solutions [10,11,12,13]. Upon the variation of solution's composition, peaks in viscosity are often observed [14-19], which can be explained with the formation of giant entangled wormlike micelles and their size and shape transformations [9,20,21,22]. The prediction and control of micelle growth and viscosity of formulations from concentrated surfactant solutions is an issue of primary importance for the personal-care and house-hold detergency [23-26], as well as in oilfield industry [27]. A comprehensive review on wormlike micelles can be found in our recent article, Ref. [28]. Here, we focus our attention on studies that are related to the subject of the present article, viz. how does the mixing of surfactant chains of different length in the micellar core affect the growth of wormlike micelles.

The addition of fatty acids of different chainlength to surfactant solutions was found to induce a significant rise of viscosity and micelle growth [18,29,30,31]. At that, the greater the *mismatch* between the chainlengths of the fatty acid and the basic surfactants, the stronger is the viscosity rise [29-31]. This fact indicates that the micelle growth is promoted by a *nonideal mixing* of the surfactant chains in the micelle hydrocarbon core and implies that the *chain-conformation* component of micelle free energy plays an important role for the formation of giant micellar aggregates. Hence, the quantitative theoretical prediction of the conformational free energy, $F_{conf}$, represents a central problem in the theory of micelle growth.

The first molecular-thermodynamic approach to the calculation of $F_{conf}$ was proposed in the pioneering works of Dill and Flory [32,33,34] on the molecular conformations in surfactant micelles. Based on calculations of the probability distribution function of chain segments in micellar aggregates, Ben-Shaul and coauthors [35,36,37] developed a model for the conformational free energy of the single-component surfactant micelles, which was extended to mixed micelles and bilayers [38,39]; for review, see Ref. [40]. Their approach was applied to model the growth of single-component nonionic surfactant micelles [41] and mixed micelles from alkyl poly(ethylene oxide) surfactants [42,43]. The complexity of the respective numerical calculations of $F_{conf}$, as well as the assumption for ideal mixing [43], does not facilitate the application of this theory works to the case of nonideal chain mixing.

An alternative approach is based on the relatively simple analytical expressions for the chain conformation free energy derived by Semenov [44] in the framework of a mean-field theory. Combining these expressions, Nagarajan and Ruckenstein [45] obtained simple and convenient analytical formulas for the conformational free energy in the case of spherical, cylindrical, and lamellar micelles. Based on these formulas, Nagarajan [46,47] proposed also a semi-empirical expression for $F_{conf}$ for binary mixed micelles. In the case of single-component ionic surfactant micelles, Kshevetskiy and Shchekin [48] found that the conformational free energy is dominated by the electrostatic and other contributions to the work of micellization. However, (as already mentioned) in the case of mixed micelles with chainlength mismatch, the experiments [29-31] indicate that the chain-conformation effects are of primary importance.

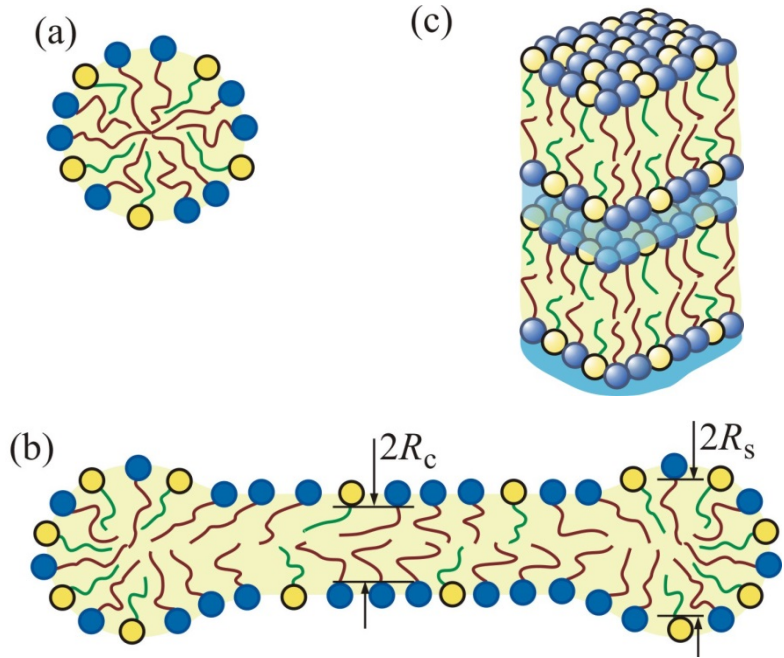

**Fig. 1**. Schematic representation of the structure of two-component surfactant aggregates in aqueous solutions: (a) spherical aggregate; (b) spherocylindrical (rodlike, wormlike) micelle; $R_c$ and $R_s$ are the radii of the cylindrical part and spherical endcaps, and (c) two neighboring lamellar aggregates.

It should be noted that Semenov [44] did not publish the full mathematical derivation of his formulas for the chain-conformation free energy of single-component aggregates. This was an obstacle to the generalization of his theory to mixed micelles. In a recent study, we succeeded to reproduce his derivations and obtained a relatively simple generalized formula for the chain-conformation free energy per molecule, $f_{conf}$, in a single-component micelle [28]:

$$\frac{f_{conf}}{k_B T} \equiv \frac{F_{conf}}{k_B T N} = \frac{3\pi^2 R^2}{16 l_{sg} l} \frac{4p^2}{1+3p+2p^2} \tag{1.1}$$

Here, $N$ is the micelle aggregation number; $l$ is the length of the extended surfactant hydrocarbon chain; $l_{sg}$ is the length of a segment in the chain; $p = V/(SR)$ is the packing parameter, where $R$, $S$ and $V$ are the radius, surface area and volume of the micelle hydrocarbon core. For spherical, cylindrical and lamellar micelles (Fig. 1), $p$ takes values 1/3, 1/2 and 1, respectively. In Ref. [28], it was demonstrated that in combination with the other components of the micelle free energy, Eq. (1.1) provides an excellent theoretical description of the growth of single-component wormlike micelles from nonionic surfactants (polyoxyethylene alkyl ethers).

Our goal in the present article is to generalize Eq. (1.1) for the case of binary mixed micelles and to give a quantitative theoretical description of the nonideal mixing of chains of different lengths. For this goal, in Section 2 we introduce the chain distribution functions for the case of mixed micelles. Section 3 is dedicated to the procedure for determining the explicit form of these functions by minimization of the conformational free energy. In Section 4, the chain-end distribution functions of the two components are determined and a formula that generalizes Eq. (1.1) is derived. Numerical results are presented in Section 5. In a subsequent article [49], the expression for $f_{conf}$ derived here is combined with the other components of micelle free energy to achieve an excellent agreement with experimental data for mixed nonionic micelles without using any adjustable parameters. Our plan is to extend this study also to mixed micelles containing ionic and zwitterionic surfactants.

## 2. Theoretical description of surfactant chain conformations

### 2.1. Chain-end distribution functions

To generalize the theory for single component micelles [28,44] to the case of two components, let us consider the surfactant chains as continuous strings with extended length $l_k = N_{sg,k} l_{sg}$ and volume $v_k = N_{sg,k} l_{sg}^3 = l_k l_{sg}^2$, where $N_{sg,k}$ is the number of segments with characteristic length $l_{sg}$; the index $k$ numbers the two surfactant components ($k = 1,2$); see Fig. 2. By definition, we assume that component 1 has longer chain, i.e. $l_1 \geq l_2$. The radial axis $r$ is directed from the surface of micelle hydrocarbon core, where $r = 0$, toward the micelle center, where $r = R$. For lamellar micelles, $r = R$ corresponds to the micelle midplane; see Fig. 1c. The outer end of the chain is anchored to the micelle surface, whereas the inner end of the chain is located in the micelle interior, at $r = r_{0k}$.

The distribution of the chain free ends inside the micelle is characterized by the function $G_k(r_{0k})$. By definition, $G_k(r_{0k})dr_{0k}$ gives the number of chains of component $k$, whose ends are located in the interval $(r_{0k}, r_{0k}+dr_{0k})$. The integration of $G_k(r_{0k})$, which is equivalent to summation over all molecules of component $k$ in the micelle, yields:

$$\int_0^R G_k(r_{0k})\,dr_{0k} = N_k \quad (k=1,2) \tag{2.1}$$

where $N_k$ is the total number of molecules of component $k$ in the micelle.

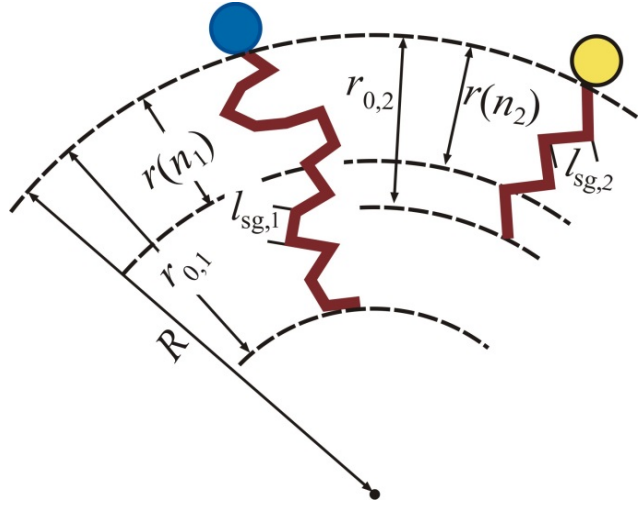

**Fig. 2**. Each surfactant chain is anchored at the micelle surface ($r = 0$) and its end is at $r = r_{0,k}$ for a molecule of the the $k$th component, $k = 1,2$; $n_k$ numbers the segments in the chain of the respective molecule; $r(n_k)$ is the distance from the $n_k$-th segment to the micelle surface; $l_{sg}$ is the length of a segment. By definition, it is assumed that component 2 has shorter chain than component 1.

*2.2. Chain-extension distribution functions*

The shape of the chain of a surfactant molecule from component $k$ within the micelle can be characterized by the function $r(n_k)$, where $n_k$ is the number of the segment, $1 \le n_k \le N_{sg,k}$, and $r$ is its radial distance from the micelle surface. By definition, at the micelle surface we have $r(0) = 0$, whereas $r(N_{sg,k}) = r_{0k}$ for the inner end of the chain. Furthermore, the local extension of the chain (in radial direction) can be characterized by the function [28,44]:

$$E_k(r_{0k},r) = \frac{dr}{dn_k}, \quad 0 < r_{0k} \le R \tag{2.2}$$

The maximal value of this distribution function of local chain extension is $E_k(r_{0k},r) = l_{sg}$ for a segment that is oriented in radial direction ($dr = l_{sg}$, $dn_k = 1$). The reciprocal quantity, $1/E_k(r_{0k},r) = dn_k/dr$, expresses the number of segments (from a given molecule of component $k$) per unit length in radial direction. The integral of this quantity gives the total number of segments in the chain:

$$\int_0^{r_{0k}} \frac{dr}{E_k(r_{0k},r)} = N_{sg,k}, \quad 0 < r_{0k} \leq R \quad (k = 1, 2) \tag{2.3}$$

Furthermore, let $S(r)$ be the area of a surface $r$ = const. situated in the micelle core (Fig. 2). The number of segments of chains contained in the elementary volume $S(r)dr$ can be presented in the form:

$$\frac{S(r)dr}{l_{sg}^3} = \int_r^R (dn_1) G_1(r_{01}) \, dr_{01} + \int_r^R (dn_2) G_2(r_{02}) \, dr_{02} \tag{2.4}$$

Here, $l_{sg}^3$ is the volume per segment; $(dn_k)$ is the number of segments (from a given molecule of component $k$) located in the considered layer of thickness $dr$ and, finally, the integration with respect to $r_{0k}$ is equivalent to summation over all surfactant molecules of component $k$, whose chain-ends are located in the interval $(r, R)$ and which contribute to the total number of segments contained in the elementary volume $S(r)dr$. It has been assumed that the segment density in the micellar core is uniform. In view of Eq. (2.2), we can represent Eq. (2.4) in the form:

$$S(r) = S_1(r) + S_2(r), \text{ where } S_k(r) = l_{sg}^3 \int_r^R \frac{G_k(r_{0k})}{E_k(r_{0k},r)} dr_{0k} \quad (k = 1, 2) \tag{2.5}$$

where $S(r)$ is the area of the surface $r$ = const., which is located at a distance $r$ from the respective interface.

*2.3. Expression for the cross-sectional area function*

The standard definition of the packing parameter, $p$, is [50]:

$$p \equiv \frac{V}{S(0)R} \tag{2.6}$$

where $S(0)$ is the surface (at $r = 0$) area of the micellar hydrocarbon core; $R$ and $V$ are the radius and volume of micelle hydrocarbon core; see Fig. 2. (In the case of lamella, $R$ and $V$ refer to the thickness and volume of one leaflet of the lamella.) For spherical, cylindrical and lamellar geometries, we have [28]:

$$V = \frac{4\pi}{3}R^3, \quad S(r) = 4\pi(R-r)^2, \quad p = \frac{1}{3} \quad \text{(sphere)} \tag{2.7}$$

$$V = \pi R^2 L_c, \quad S(r) = 2\pi(R-r)L_c, \quad p = \frac{1}{2} \quad \text{(cylinder)} \tag{2.8}$$

$$V = RS_{\text{lam}}, \quad S(r) = S_{\text{lam}}, \quad p = 1 \quad \text{(lamella)} \tag{2.9}$$

where $0 \leq r \leq R$; $L_c$ is the length of the cylinder, and $S_{\text{lam}}$ is the area of the lamella. The above expressions for $S(r)$ can be presented in a compact form [28]:

$$S(r) = S(0)(1-r/R)^{\frac{1-p}{p}} = \frac{V}{Rp}(1-r/R)^{\frac{1-p}{p}} \tag{2.10}$$

One could verify that for $p = 1/3$, $1/2$ and $1$ Eq. (2.10) reduces to the expression for $S(r)$ in Eqs. (2.7), (2.8) and (2.9), respectively. Following Ref. [28], we assume that Eq. (2.10) can be used as an interpolation formula for $S(r)$ in the whole interval $1/3 \leq p \leq 1$.

## 3. Chain-conformation free energy

### 3.1. Expression for the conformation free energy

The unperturbed end-to-end distance of a chain of component $k$ containing $dn_k$ segments is $(dn_k)^{1/2} l_{\text{sg}}$ [51]. Inside the micelle, this chain could be extended, so that its ends lie at a distance $dr$ from one another. This corresponds to a local conformation elastic free energy of the considered chain, $\varphi_{\text{conf},k}$, which can be estimated by using the Flory expression [44,45,51]:

$$\frac{d\varphi_{\text{conf},k}}{k_B T} = \frac{3}{2} \frac{(dr)^2}{[(dn_k)^{1/2} l_{\text{sg}}]^2} = \frac{3}{2 l_{\text{sg}}^2} E_k(r_{0k}, r) dr \tag{3.1}$$

The factor 3/2 accounts for the three-dimensional character of deformation; $k_B$ is the Boltzmann constant and $T$ is the temperature. By integration of Eq. (3.1) and summation over all components, one obtains the total conformational free energy for all chains in the micellar core:

$$\frac{F_{\text{conf}}}{k_B T} = \frac{3}{2 l_{\text{sg}}^2} \sum_{k=1}^{2} \int_0^R dr_{0k} G_k(r_{0k}) \int_0^{r_{0k}} dr E_k(r_{0k}, r) \tag{3.2}$$

In the last equation, the integration over $r$ is equivalent to summation of contributions from all segments of a given molecule, whereas the integration over $r_{0k}$ – to summation over all molecules of component $k$. Furthermore, it is convenient to introduce dimensionless variables as follows:

$$x_k = \frac{r_{0k}}{R}, \quad y = \frac{r}{R}, \quad \varepsilon_k(x_k, y) = \frac{N_{sg,k}}{R} E_k(r_{0k}, r), \quad g_k(x_k) = \frac{Rv_k}{V} G_k(r_{0k}) \quad (k = 1, 2) \tag{3.3}$$

In terms of the new variables, Eqs. (2.1) and (2.3) acquire the form:

$$\int_0^1 g_k(x_k)\, dx_k = \eta_k \quad (k = 1, 2) \tag{3.4}$$

$$\int_0^{x_k} \frac{dy}{\varepsilon_k(x_k, y)} = 1 \quad \text{for } 0 < x_k \leq 1 \quad (k = 1, 2) \tag{3.5}$$

where $\eta_1$ and $\eta_2$ are the *volume* fractions of the tails of the two surfactants in the micellar core:

$$\eta_k \equiv \frac{N_k v_k}{V} \quad (k = 1, 2); \quad \eta_1 + \eta_2 = 1 \tag{3.6}$$

It is convenient to introduce also local *area* fractions, $s_1(y)$ and $s_2(y)$, occupied by the tails of the two surfactant components in a cross-section corresponding to a dimensionless distance $y = r/R$:

$$s_k(y) \equiv \frac{S_k(r)}{S(r)} \quad (k = 1, 2); \quad s_1(y) + s_2(y) = 1 \tag{3.7}$$

With the help of Eqs. (2.10), (3.3) and (3.7), we can represent Eq. (2.5) in the form

$$\frac{s_k(y)}{p}(1-y)^{\frac{1-p}{p}} = \int_y^1 \frac{g_k(x_k)}{\varepsilon_k(x_k, y)}\, dx_k \quad \text{for } 0 \leq y \leq 1 \quad (k = 1, 2) \tag{3.8}$$

Likewise, in terms of the dimensionless variables Eq. (3.2) acquires the form:

$$f_{conf} \equiv \frac{F_{conf}}{k_B T N} = \frac{3R^2}{2l_{sg}} \sum_{k=1}^{2} \frac{y_k}{l_k \eta_k} \int_0^1 dx_k g_k(x_k) \int_0^{x_k} dy\, \varepsilon_k(x_k, y) \tag{3.9}$$

where $f_{conf}$ is the dimensionless conformation free energy per molecule; $y_1$ and $y_2$ are the *mole* fractions of the two surfactant components in the micelle:

$$y_k \equiv \frac{N_k}{N} \quad (k = 1, 2); \quad N \equiv N_1 + N_2 \text{ and } y_1 + y_2 = 1 \tag{3.10}$$

Using Eqs. (3.6) and (3.10) along with the definition $v_k = l_k l_{sg}^2$, we derive the relation between the mole and volume fractions, $y_k$ and $\eta_k$:

$$y_k = \frac{N_k}{N} = \frac{1}{N}\frac{\eta_k V}{v_k} = \frac{\bar{l}}{l_k}\eta_k \quad (k=1,2); \quad \bar{l} \equiv y_1 l_1 + y_2 l_2 \tag{3.11}$$

where the micelle core volume is $V = v_1 N_1 + v_2 N_2$, and $\bar{l}$ is an average chainlength. Substituting $y_k$ from Eq. (3.11) into Eq. (3.9), we finally obtain:

$$\frac{f_{\text{conf}}}{k_B T} = \frac{3R^2}{2l_{sg}}\sum_{k=1}^{2}\frac{\bar{l}}{l_k^2}\int_0^1 dx_k\, g_k(x_k)\int_0^{x_k} dy\, \varepsilon_k(x_k, y) \tag{3.12}$$

*3.2. Formulation of the minimization problem*

The variational problem reduces to minimization of the free energy functional in Eq. (3.12) with respect to variations of four functions, viz. $g_k(x_k)$ and $\varepsilon_k(x_k,y)$ for $k = 1,2$, which describe the conformations of surfactant chains in the micelle. The minimization is to be carried out in the presence of four constraints expressed by Eqs. (3.5) and (3.8) for $k = 1,2$. As explained in Appendix A, Eq. (3.4) does not lead to additional constraints.

To solve the minimization problem, we introduce Lagrange multipliers $\lambda_k(x_k)/l_k^2$ and $\gamma_k(y)/l_k^2$, which are associated with the independent constraints expressed by Eqs. (3.5) and (3.8), respectively. Thus, in view of Eq. (3.12) the problem is reduced to minimization of the following Lagrange functional [28]

$$\Phi \equiv \sum_{k=1}^{2}\frac{1}{l_k^2}\int_0^1 dx_k\, g_k(x_k)\int_0^{x_k} dy\, \varepsilon_k(x_k,y) - \sum_{k=1}^{2}\frac{1}{l_k^2}\int_0^1 dx_k\, \lambda_k(x_k)\left[\int_0^{x_k}\frac{dy}{\varepsilon_k(x_k,y)} - 1\right]$$

$$- \sum_{k=1}^{2}\frac{1}{l_k^2}\int_0^1 dy\, \gamma_k(y)\left[\int_y^1 dx_k\, \frac{g_k(x_k)}{\varepsilon_k(x_k,y)} - \frac{s_k(y)}{p}(1-y)^{\frac{1-p}{p}}\right] \tag{3.13}$$

with respect to variations of the functions $g_k(x_k)$ and $\varepsilon_k(x_k,y)$ for $k = 1,2$. The minimization is to be carried out at fixed area fractions of the chains, $s_1(y)$ and $s_2(y)$.

*3.3. Results from the minimization procedure*

The minimization problem is solved analytically in Appendix A. The following expressions for the chain distribution functions have been obtained:

$$\varepsilon_k(x_k, y) = \frac{\pi}{2}(x_k^2 - y^2)^{1/2} \quad (k=1,2) \tag{3.14}$$

$$g_k(x) = -\frac{d}{dx}[\int_x^1 \frac{s_k(y)y(1-y)^{\frac{1-p}{p}}}{p(y^2-x^2)^{1/2}} dy] \quad \text{for } 0 \le x \le 1 \quad (k=1,2) \qquad (3.15)$$

In view of Eq. (3.7), the summation in Eq. (3.15) yields:

$$g(x) \equiv g_1(x) + g_2(x) = -\frac{d}{dx}[\int_x^1 \frac{y(1-y)^{\frac{1-p}{p}}}{p(y^2-x^2)^{1/2}} dy] \quad \text{for } 0 \le x \le 1 \qquad (3.16)$$

The last expression represents exactly the chain-end distribution function, $g(x)$, in the case of single-component surfactant micelles [28]. In the case of binary surfactant mixtures $g(x) = g_1(x) + g_2(x)$ plays the role of *total* chain-end distribution function. As demonstrated in Section 4.1, the chain-distribution functions of the separate components, $g_1(x)$ and $g_2(x)$, can be expressed in terms of $g(x)$.

For numerical calculation of $g(x)$, it is convenient to use the substitution $y = x + t^2$ and to represent Eq. (3.16) in the following equivalent form:

$$g(x) = 2\frac{1-p}{p^2} x \int_0^{(1-x)^{1/2}} \frac{(1-x-t^2)^{\frac{1-2p}{p}}}{(2x+t^2)^{1/2}} dt \qquad (3.17)$$

*3.4. The total chain-end distribution function $g(x)$*

In the special cases of spherical ($p = 1/3$), cylindrical ($p = 1/2$) and lamellar ($p = 1$) micelles, the integral in Eq. (3.17) can be solved in terms of elementary functions. The respective expressions for $g(x)$ read [28,44]:

$$g(x) = 6x\left\{\ln\left[\frac{1}{x} + (\frac{1}{x^2}-1)^{1/2}\right] - (1-x^2)^{1/2}\right\} \quad \text{(sphere)} \qquad (3.18)$$

$$g(x) = 2x\ln[\frac{1}{x} + (\frac{1}{x^2}-1)^{1/2}] \quad \text{(cylinder)} \qquad (3.19)$$

$$g(x) = \frac{x}{(1-x^2)^{1/2}} \quad \text{(lamella)} \qquad (3.20)$$

For the spherical endcaps of wormlike micelles (which have the shape of truncated spheres), the packing parameter takes values in the interval $1/3 \le p \le 3/8$ [28]. In this case, $g(x)$ can be calculated by numerical integration in Eq. (3.17) by using, e.g., the Simpson rule.

Fig. 3 shows the graphs of the function $g(x)$ for $p = 1/3, 3/8, 1/2,$ and 1. In all cases, $g(0) = 0$, which is seen also from Eq. (3.17). Physically, this means that the probability to find a chain-end at the micelle surface is negligible. For $p = 1/3$, $3/8$ and $1/2$, we have also $g(1) = 0$, so that the function $g(x)$ has a maximum at an intermediate value of $x$, which corresponds to the maximal density of surfactant chain-ends in the micelle.

In the case of lamellar micelle ($p = 1$), $g(x)$ is a monotonically increasing function that has a weak (integrable) singularity for $x \to 1$ (Fig. 3). In this case, the probability to find a surfactant chain-end in the center of the lamella (at $x = 1$) is the greatest.

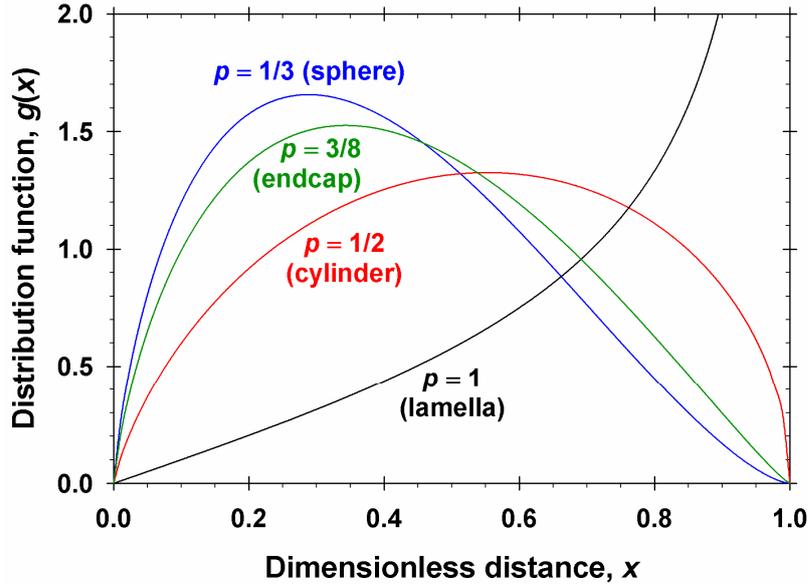

**Fig. 3.** Graphs of the total chain-end distribution function, $g$, vs. the dimensionless distance to the micelle surface, $x$, for four different values of the packing parameter $p$ corresponding to different micellar shapes.

The mathematical investigation of the function $g(x)$ defined by Eq. (3.17) gives the following results for the asymptotic behavior of $g(x)$ at $x \to 1$:

$$g(1) = \begin{cases} 0 & \text{for } p < 2/3 \\ 3\pi\sqrt{2}/8 \approx 1.666 & \text{for } p = 2/3 \\ \infty & \text{for } p > 2/3 \end{cases} \quad (3.21)$$

The fact that for $0 < p < 2/3$ the chain-end function $g(x)$ has a maximum (Fig. 3) is related to the specific geometry of the micelles in this range of $p$ values, viz. that the center of curvature of the micelle surface is located inside the micelle, as it is for the spherical and cylindrical micelles, and for the endcaps, as well. Because, the hydrophobic chains of the surfactant molecules have finite volume, only the ends of a few chains could be located in the small vicinity of the center of curvature. Mathematically, in the framework of the used mean-field approach, this leads to $g(x) \to 0$ for $x \to 1$. In addition, $g(x) = 0$ for $x \to 0$ (the chains begin, rather than end, at the micelle surface). In such a case, because in general $g(x) > 0$ in the micelle interior ($0 < x < 1$), the known Rolle's theorem implies that the function $g(x)$ must have a maximum for $0 < x < 1$. With the rise of $p$, say, if we compare cylinder with sphere, the space in the vicinity of micelle center increases, which allows the ends of more molecules to enter this vicinity, and for this reason the position of the maximum moves to the right in Fig. 3.

**4. Distributions of the shorter and longer chains**

*4.1. Determination of $g_1(x)$ and $g_2(x)$*

First, in Eq. (3.12) we substitute $\varepsilon_k(x_k,y)$ from Eq. (3.14) and set $g_1(x) = g(x) - g_2(x)$. After some transformations, we obtain:

$$\frac{f_{\text{conf}}}{k_B T} = \frac{3\pi^2 R^2 \bar{l}}{16 l_{sg}} \left[ \frac{1}{l_1^2} \int_0^1 x^2 g(x) \, dx + (\frac{1}{l_2^2} - \frac{1}{l_1^2}) \int_0^1 x^2 g_2(x) \, dx \right] \qquad (4.1)$$

If the two surfactants have equal chainlengths, $l_1 = l_2 = l$, then Eq. (4.17) reduces to Eq. (1.1); see Ref. [28] for details.

In the general case $l_1 \neq l_2$, in view of Eq. (3.15) the distribution function of the shorter chains, $g_2(x)$, depends on the unknown function $s_2(y)$. [If $s_2(y)$ and $g_2(x)$ are determined, we could find also $s_1(y) = 1 - s_2(y)$ and $g_1(x) = g(x) - g_2(x)$.] To determine $g_2(x)$, we subject the configurational free energy, $f_{\text{conf}}$, to an additional minimization with respect to the variations of $g_2(x)$. Because $g_2(x)$ appears only in the last integral in Eq. (4.1) and $l_2 < l_1$ (by definition), the minimization of $f_{\text{conf}}$ is equivalent to minimization of the following functional:

$$\Lambda[g_2] = \int_0^1 x^2 g_2(x) \, dx \qquad (4.2)$$

A physical constraint on $g_2(x)$ follows from Eq. (3.16):

$$0 \leq g_2(x) \leq g(x) \text{ for } 0 \leq x \leq 1 \qquad (4.3)$$

An additional physical constraint follows from Eq. (3.4):

$$\int_0^1 g_2(x) \, dx = \eta_2 \qquad (4.4)$$

In Appendix B it is proven that the inequality $0 \leq s_2(y) \leq 1$ [with $s_2(y)$ substituted from Eq. (3.8)] does not lead to any additional constraints on $g_2(x)$. Thus, the problem is reduced to the minimization of the functional in Eq. (4.2) under constraints given by Eqs. (4.3) and (4.4). A logical solution to the problem can be found by discretization of the integrals in Eqs. (4.2) and (4.4).

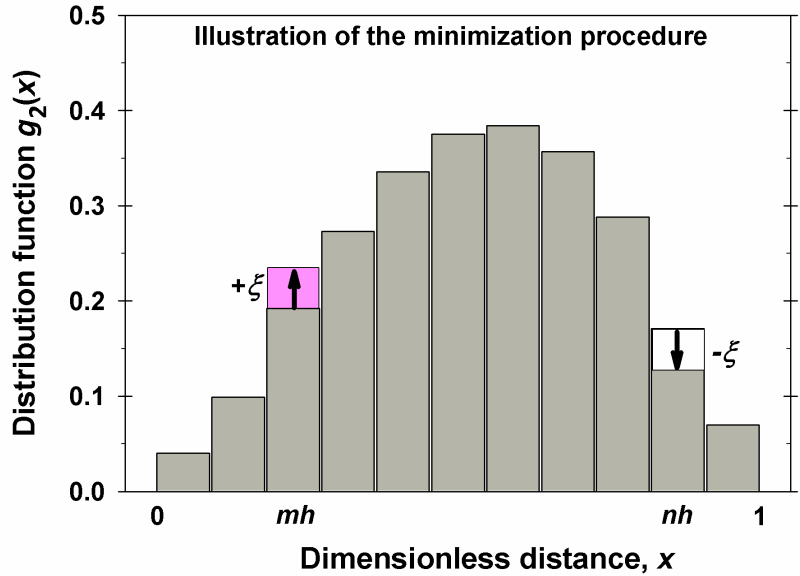

**Fig. 4.** Illustration of the discretization of the short-chain distribution function $g_2(x)$, which is used to determine $g_2(x)$ by minimization of the functional $\Lambda[g_2]$; details in the text.

For this goal, the interval [0, 1] is divided to $N_s$ identical segments of length $h$, and the trapezoidal rule is applied. Thus, the discretized form of Eqs. (4.4) and (4.2) reads:

$$\sum_{j=1}^{N_s-1} g_2(jh)h + \frac{g_2(1)}{2}h = \eta_2 \qquad (4.5)$$

$$\Lambda[g_2] = \sum_{j=1}^{N_s-1} j^2 h^2 g_2(jh)h + \frac{g_2(1)}{2}h \qquad (4.6)$$

where we have taken into account that $g_2(0) = 0$. Furthermore, let us imagine a variation of $g_2$, which consists of small increase of $g_2$ with an increment $\xi$ at $j = m$ (Fig. 4). To have a constant value in the right hand side of Eq. (4.5), the same variation should include also a decrease of $g_2$ with $\xi$ at another value of the summation index, $j = n$ (Fig. 4). At that, the value of $\Lambda[g_2]$ in Eq. (4.6) becomes:

$$\Lambda[g_2] = \sum_{j=1}^{N_s-1} j^2 h^3 g_2(jh) + \frac{g_2(1)}{2}h + (m^2 - n^2)h^3\xi \qquad (4.7)$$

By definition, $m < n$. Then, if the considered variation leads to a *smaller* value of $\Lambda$, we should have $\xi > 0$. At the minimum of $\Lambda[g_2]$, for the smaller values of $j$ (like $j = m$) the function $g_2$ should reach its greatest possible value, $g_2(x) = g(x)$, whereas for larger values of $j$ (like $n$) $g_2$ should reach its smallest possible value, $g_2(x) = 0$; see Eq. (4.3). Thus, following the logic of a numerical minimization we arrive to the conclusion that the function, which minimizes the functional $\Lambda[g_2]$, should be (Fig. 5)

$$g_2(x) = \begin{cases} g(x) & \text{for } 0 \leq x \leq b \\ 0 & \text{for } b < x \leq 1 \end{cases} \tag{4.8}$$

where $g(x)$ is given by Eq. (3.17) and $b$ is the coordinate of a transition point, which is defined in such a way that the constraint given by Eq. (4.4) is satisfied, viz.:

$$\int_0^b g(x)\,dx = \eta_2 \tag{4.9}$$

In view of the relation $g_1(x) = g(x) - g_2(x)$ and Eq. (4.8), the distribution function of the ends of the longer chains is

$$g_1(x) = \begin{cases} 0 & \text{for } 0 \leq x < b \\ g(x) & \text{for } b \leq x \leq 1 \end{cases} \tag{4.10}$$

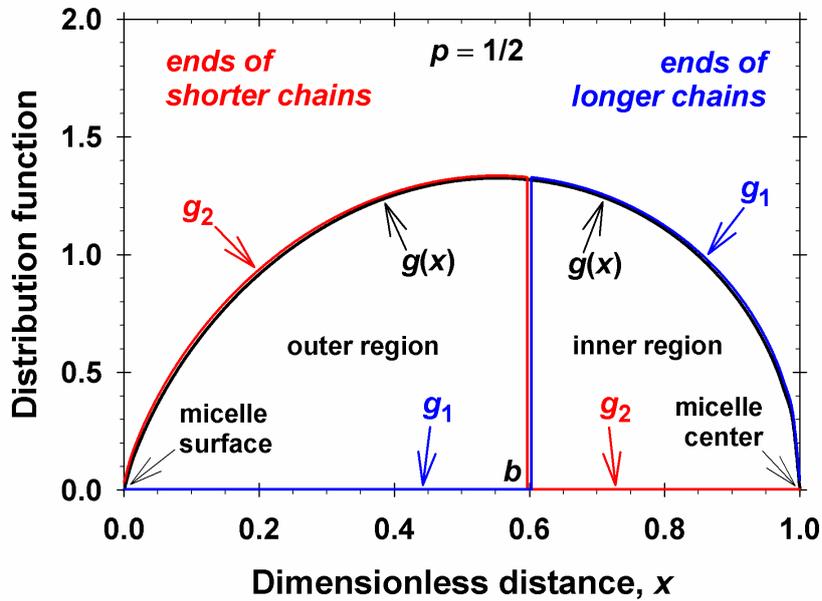

**Fig. 5.** Distribution functions of the ends of the longer and shorter chains, $g_1(x)$ and $g_2(x)$, respectively, as predicted by Eqs. (4.8) and (4.10) – illustration for the case of cylindrical micelle, $p = 1/2$; $x = b$ is the boundary between the regions, where the ends of the shorter and longer chains are located; $g(x)$ is the distribution function in the case of identical chains.

The graphs of the functions $g_1(x)$ and $g_2(x)$ defined by Eqs. (4.8) and (4.10) are shown in Fig. 5 for the special case of cylindrical micelle, $p = 1/2$. For $0 < x < b$, the distribution of the ends of the shorter chains, $g_2(x)$, coincides with the chain-end distribution $g(x)$ for single-component micelles, whereas no ends of the longer chains are located in this zone, where $g_1(x) = 0$. In other words, the chains of the longer surfactant molecules lie across the *outer* region, $0 < x < b$, their ends being located in the *inner* region of the micellar core, $b < x < 1$. In contrast, the shorter chains do not penetrate in the inner region, where $g_2(x) = 0$.

Substituting $\varepsilon_k(x_k,y)$ and $g_k(x)$ from Eqs. (3.14), (4.8) and (4.10) into Eq. (3.8), one could calculate also the functions $s_k(y)$, $k = 1,2$.

*4.2. Determination of the boundary between the outer and inner regions*

To calculate the parameter $b$, which determines the boundary between the outer and inner regions of the micellar core, it is more convenient to use the relation

$$\int_b^1 g(x)\,dx = \eta_1 \tag{4.11}$$

which is a corollary from Eqs. (3.4) and (4.10) and is mathematically equivalent to Eq. (4.9) in view of the identities $g_1 + g_2 = g$ and $\eta_1 + \eta_2 = 1$. The substitution of $g(x)$ from Eq. (3.16) into Eq. (4.11) after some transformations yields:

$$\int_b^1 \frac{y(1-y)^{\frac{1-p}{p}}}{p(y^2 - b^2)^{1/2}}\,dy = \eta_1 \tag{4.12}$$

To remove the singularity of the integrand at $y = b$, it is convenient to use the substitution $y = b + t^2$, which brings Eq. (4.12) in the form:

$$\frac{2}{p}\int_0^{(1-b)^{1/2}} \frac{(b+t^2)(1-b-t^2)^{\frac{1-p}{p}}}{(2b+t^2)^{1/2}}\,dt = \eta_1 \tag{4.13}$$

For given $\eta_1$ and $p$, the coordinate of the boundary point, $b$, can be calculated by numerical integration in Eq. (4.13) by using, e.g., the Simpson rule. In the special cases of sphere ($p = 1/3$), cylinder ($p = 1/2$) and lamella ($p = 1$), the integral in Eq. (4.12) can be taken analytically, which leads to the following equations for $b$:

$$(1-b^2)^{1/2}(1+2b^2) - 3b^2 \ln[\frac{1}{b} + (\frac{1}{b^2} - 1)^{1/2}] = \eta_1 \quad \text{(sphere)} \tag{4.14a}$$

$$(1-b^2)^{1/2} - b^2 \ln\left[\frac{1}{b} + \left(\frac{1}{b^2}-1\right)^{1/2}\right] = \eta_1 \quad \text{(cylinder)} \tag{4.14b}$$

$$b = (1-\eta_1^2)^{1/2} \quad \text{(lamella)} \tag{4.14c}$$

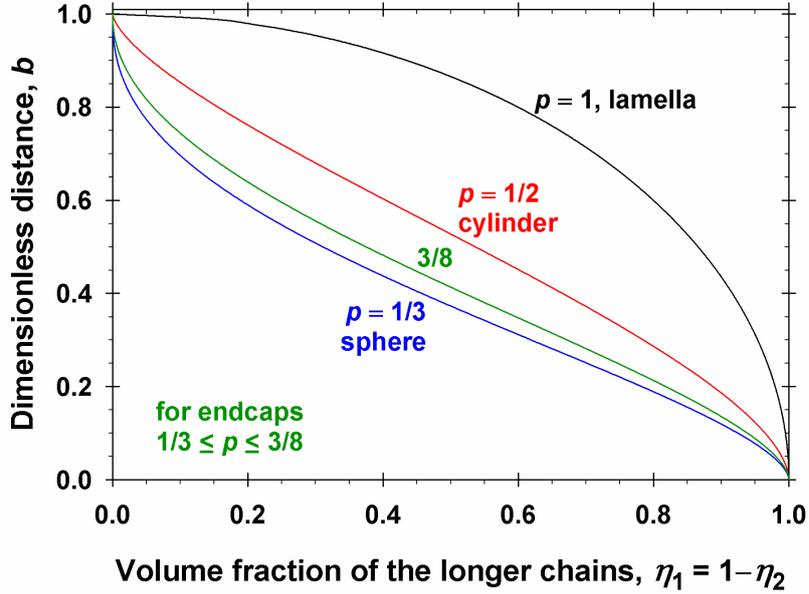

**Fig. 6.** Position, $b$, of the boundary between the regions, where the ends of the shorter and longer chains are located, plotted vs. the volume fraction of the longer chains, $\eta_1$, for four different values of the packing parameter, $p$, corresponding to different micellar shapes.

Fig. 6 shows plots of $b$ vs. $\eta_1$ calculated from Eq. (4.13) or (4.14) for $p = 1/3$, 3/8, 1/2, and 1. One sees that $b$ (and the thickness of the outer region) increases with the rise of $p$. Moreover, $b \to 0$ for $\eta_1 \to 1$, that is the region of the shorter chains ($0 < x < b$) vanishes and only the longer chains remain; see Fig. 5.

In the other limit we obtain $b \to 1$ for $\eta_1 \to 0$ (Fig. 6), i.e. the region where the ends of the longer chains are located ($b < x < 1$) vanishes and only the shorter chains remain (Fig. 5).

Because of the relation $\eta_1 = 1 - \eta_2$, Fig. 6 gives the dependence of the boundary coordinate, $b$, also on the volume fraction of the shorter chains, $\eta_2$. As demonstrated in this figure, the solution of Eq. (4.13) [or of its special forms, Eqs. (4.14a) – (4.14c)] for $b$ always exists for any given $\eta_1$ ($0 \le \eta_1 \le 1$). Physically, this means that if the two surfactants have different chainlengths, $l_2 \ne l_1$, the inner and outer domains shown in Fig. 5 always exist, and that the central part of the micelle ($b < x < 1$) is always occupied only by the longer chains.

It should be noted that Fig. 5 illustrates the positions of the outer and inner regions for the special case of cylindrical micelle ($p = 1/2$). For other values of the packing parameter, e.g. for $p = 1/3$, $3/8$ and 1, the positions of the outer and inner regions of micelle core can be determined by simply taking the value of $b$ for the respective $\eta_1$ and $p$ in Fig. 6, and then, by plotting the vertical boundary line $x = b$ for the $g(x)$ curve with the respective $p$-value in Fig. 3.

It should be also noted that the ratio of chainlengths of the two surfactants, $l_1/l_2$, affects the graph in Fig. 5 through the value of $b$ that, in turns, depends on $\eta_1 = 1-\eta_2$ (Fig. 6), which is related to $l_1/l_2$ in view of Eq. (3.11):

$$\frac{\eta_1}{\eta_2} = \frac{l_1}{l_2}\frac{y_1}{y_2} \tag{4.15}$$

As usual, $y_k$ and $\eta_k$ ($k = 1,2$) are the molar and volume fractions of the chains, respectively.

As an example, taking the value $b = 0.6$ in Fig. 5, from Eq. (4.14b) we obtain $\eta_1 \approx 0.4045$, which means that $\eta_2 \approx 0.5955$. Furthermore, the lengths of $C_{14}$ and $C_{10}$ alkyl chains are $l_1 = 1.92$ nm and $l_2 = 1.42$ nm; see Section 5.1. Then, from Eq. (4.15) we determine $y_1 \approx 0.33$ and $y_2 \approx 0.67$.

*4.3. Interaction parameter for two-component micelles*

In the case of *single-component* surfactant micelles, the chain-conformation free energy per molecule in the micelle is [28]:

$$\frac{f_{\text{conf}}}{k_B T} = \frac{3\pi^2 R^2}{16 l_{\text{sg}} l} c_{\text{conf}}(p) \tag{4.16a}$$

$$c_{\text{conf}}(p) \equiv \frac{2}{p}\int_0^1 z^2(1-z)^{\frac{1-p}{p}}\,\mathrm{d}z = \frac{4p^2}{1+3p+2p^2} \tag{4.16b}$$

In the case of *two-component* micelles, combining Eqs. (4.1) and (4.8) we obtain:

$$f_{\text{conf}} = \frac{3\pi^2 R^2 \bar{l}}{16 l_{\text{sg}}}[\frac{1}{l_1^2}\int_0^1 g(x)x^2\,\mathrm{d}x + (\frac{1}{l_2^2}-\frac{1}{l_1^2})\int_0^b g(x)x^2\,\mathrm{d}x] \tag{4.17}$$

where $g(x)$ and $b$ are determined by Eqs. (3.16) and (4.13), respectively. If the two surfactants have equal chainlengths, $l_1 = l_2 = l$, then Eq. (4.17) reduces to Eq. (4.16a).

For the general case of different chainlengths, $l_1 \neq l_2$, in Appendix B it is shown that the substitution of $g(x)$ from Eq. (3.16) in Eq. (4.17) leads to

$$\frac{f_{conf}}{k_B T} = \frac{3\pi^2 R^2}{16 l_{sg}} c_{conf}(p) \left[ \frac{y_1}{l_1} + \frac{y_2}{l_2} - (\frac{\overline{l}}{l_2^2} - \frac{\overline{l}}{l_1^2}) \beta_{conf}(p,\eta_1) \right] \quad (l_2 \leq l_1) \tag{4.18}$$

As before, $y_1$ and $y_2$ are the mole fractions of the two components in the micelle, and the *chain-conformation interaction parameter*, $\beta_{conf}$, is defined as follows:

$$\beta_{conf}(p,\eta_1) \equiv \frac{1}{c_{conf}(p)} \left\{ [b^2 - c_{conf}(p)]\eta_1 + \frac{2}{p} \int_b^1 z(z^2 - b^2)^{1/2} (1-z)^{\frac{1-p}{p}} dz \right\} \tag{4.19}$$

In the limiting cases $\eta_1 = 0$ and $\eta_1 = 1$, from Eq. (4.19) we obtain

$$\beta_{conf}(p,0) = \beta_{conf}(p,1) = 0 \tag{4.20}$$

i.e. $\beta_{conf}$ is zero for single-component micelles. Indeed, for $\eta_1 = 0$ we have $b = 1$ (Fig. 6) and the integral in Eq. (4.19) is zero because of the vanishing integration domain. In the other limit, for $\eta_1 = 1$ we have $b = 0$ (Fig. 6) and the integral in Eq. (4.19) becomes equal to $c_{conf}(p)$; see Eq. (4.16b).

If the interaction parameter was identically zero, $\beta_{conf}(p,\eta_1) \equiv 0$, then Eq. (4.18) would describe *ideal mixing*, viz.:

$$f_{conf} = y_1 f_{conf,1} + y_2 f_{conf,2}; \quad \frac{f_{conf,k}}{k_B T} = \frac{3\pi^2 R^2}{16 l_{sg} l_k} c_{conf}(p), \quad k = 1,2; \tag{4.21}$$

see also Eq. (4.16a). In the general case, $\beta_{conf}(p,\eta_1)$ can be calculated by solving the integral in Eq. (4.19) numerically, e.g., by using the Simpson rule.

In the special cases of spherical ($p = 1/3$), cylindrical ($p = 1/2$) and lamellar ($p = 1$) micelles, the integral in Eq. (4.19) can be solved analytically; see Appendix B. The resulting expressions for $\beta_{conf}(p,\eta_1)$ read:

$$\beta_{conf}(\frac{1}{3},\eta_1) = (1-b^2)^{5/2} + \frac{5}{2} b^2 \eta_1 - \eta_1 \quad \text{(sphere)} \tag{4.22}$$

$$\beta_{conf}(\frac{1}{2},\eta_1) = (1-b^2)^{3/2} + \frac{3}{2} b^2 \eta_1 - \eta_1 \quad \text{(cylinder)} \tag{4.23}$$

$$\beta_{conf}(1,\eta_1) = \frac{\eta_1}{2}(1-\eta_1^2) \quad \text{(lamella)} \tag{4.24}$$

For the spherical endcaps of wormlike micelles, the packing parameter takes values in the interval $1/3 < p \leq 3/8$, so that in this case $\beta_{conf}(p,\eta_1)$ is to be calculated by numerical integration in Eq. (4.19).

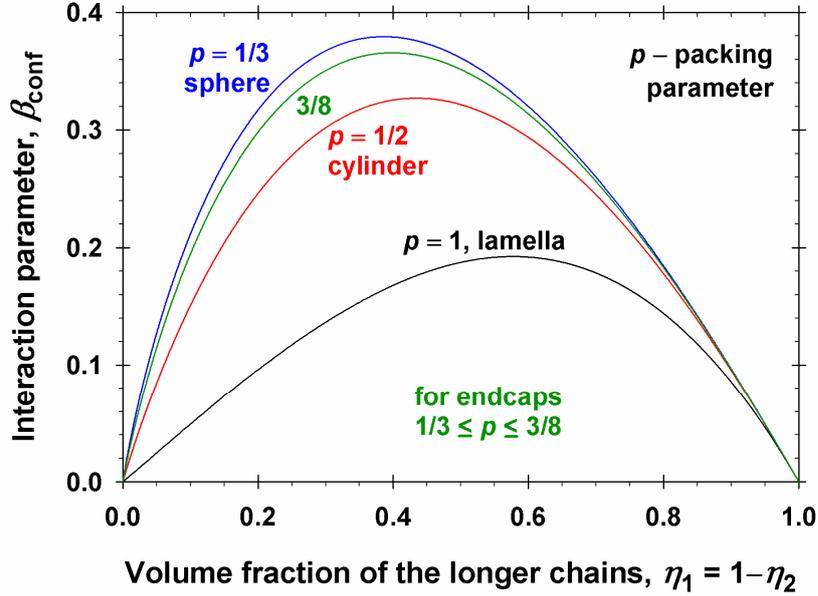

**Fig. 7.** Plot of the chain-configuration interaction parameter, $\beta_{conf}$, which accounts for the deviation from ideal mixing, vs. the volume fraction of the longer chains, $\eta_1$, for four different values of the packing parameter, $p$, corresponding to different micellar shapes.

Fig. 7 shows the graphs of $\beta_{conf}$ vs. $\eta_1$ for $p = 1/3$, $3/8$, $1/2$, and 1. Because of the relation $\eta_1 = 1 - \eta_2$, Fig. 7 gives the dependence of $\beta_{int}$ also on the volume fraction of the shorter chains, $\eta_2$. One sees that for mixed micelles with different chainlengths ($0 < \eta_1 < 1$, $l_2 < l_1$) the interaction parameter is positive, $\beta_{conf} > 0$, and then Eq. (4.18) implies that the mixing of two surfactants with different chainlengths is *always synergistic* with respect to the chain-conformation free energy $f_{conf}$. In other words, the mixing of two surfactants with different chainlengths always favors the micellization and micelle growth. Calculated coordinates of the maxima of the curves in Fig. 7 are given by Eqs. (B.11), (B.17) and (B.23) in Appendix B. These maxima correspond to the greatest deviations from ideal mixing and to a composition, which is the most favorable for micellization and micelle growth with respect to the chain conformations.

As mentioned in the Introduction, a phenomenon, which is manifestation of the nonideal chain mixing, is the experimental finding that the viscosity of surfactant solutions with wormlike micelles increases upon the addition of fatty acids, the increase being greater if the mismatch ($l_1–l_2$) between the chainlengths of surfactant and fatty acid is greater [29]. For example, in the case of added octanoic acid (shorter chain and greater mismatch) the height of viscosity peak was 41.0 Pa·s [30], vs. only 0.603 Pa·s in the case of added dodecanoic acid (longer chain and smaller mismatch) [31], all other conditions being the same.

In the literature, effects of nonideal mixing of surfactants with respect to the values of CMC and micelle growth have been discussed mostly in relation to the fact that the interactions at the *micellar surface* depend nonlinearly on the composition of the mixed surfactant systems [45,46]. What concerns the mixing of chains in the *micellar core*, Nagarajan [46,47] proposed a semiempirical formula, which assumes ideal mixing for micellar radii $R < l_1, l_2$, but nonideal mixing for $l_2 \leq R \leq l_1$; see Eqs. (B.24) and (B.25) in Appendix B and the detailed discussion therein. In the framework of the mean-field theoretical approach developed in the present study, solution of Eq. (4.13) for $b$ always exists (Fig. 6); the shorter chains are always located in the outer region ($0 \leq x \leq b$ in Fig. 5), and for $l_2 \neq l_1$ the mixing is *always* nonideal; see Eq. (4.18) and Fig. 7.

## 5. Numerical results and discussion

### 5.1. Molecular parameters

The functions $g(x)$, $g_1(x)$, $g_2(x)$, $b(p,\eta_1)$ and $\beta_{conf}(p,\eta_1)$, which are plotted in Figs. 3, 5, 6 and 7, are *universal* functions, which are independent of specific molecular parameters, e.g. of the chainlengths, $l_1$ and $l_2$, and of the length per segment, $l_{sg}$. However, to calculate the chain-conformation free energy for different surfactant mixtures, we have to specify also the values of $l_1$, $l_2$, and $l_{sg}$.

**Table 1**. Length, $l$, of surfactant alkyl chains containing $n_C$ carbon atoms.

| $n_C$ | $l$, nm |
|---|---|
| 10 | 1.42 |
| 12 | 1.67 |
| 14 | 1.92 |
| 16 | 2.18 |
| 18 | 2.43 |

For the length per segment we will use the value, $l_{sg} = 0.46$ nm, which is an appropriate value for paraffin chains suggested by Dill, Flory et al. [32-34]. The lengths of the paraffin chains can be estimated by using the Tanford formula [52]:

$$l(n_C) = l(CH_3) + (n_C - 1)l(CH_2) \tag{5.1}$$

where $n_C$ is the total number of carbon atoms in the chain; the length per CH$_3$ group is $l(CH_3) = 0.280$ nm, and the length per CH$_2$ group is $l(CH_2) = 0.1265$ nm. Table 1 presents the values of $l(n_C)$ calculated from Eq. (5.1) for even $n_C$ in the range from 10 to 18.

## 5.2. Maximal radius of the mixed micelles, $R_{max}$

By definition, $R$ denotes the radius of a spherical or cylindrical micelle, as well as the radius of the endcaps of spherocylindrical micelles. In the case of lamellar micelle, $R$ denotes its half-thickness. To determine theoretically the equilibrium value of $R$ for a micelle of given shape, its total free energy is to be minimized by variation of $R$; see e.g. [28]. To specify the interval of variation of $R$, it is important to know the maximal possible value of $R$, which will be denoted $R_{max}$. Our goal here is to determine $R_{max}$ for mixed micelles containing surfactants with two different chainlengths ($l_2 < l_1$) at various values of the packing parameter, $p$.

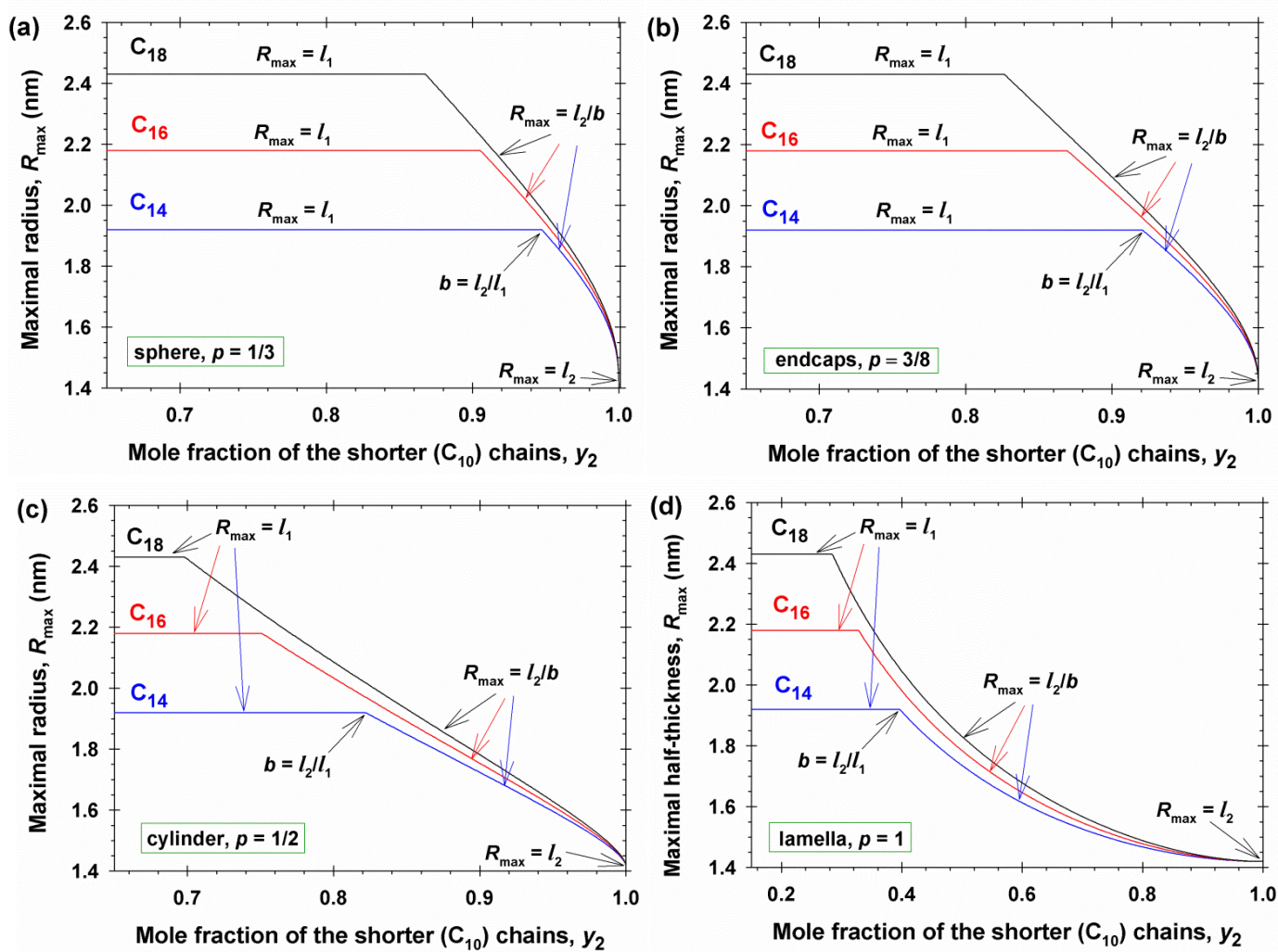

**Fig. 8**. Maximum possible radius (or half-thickness) $R_{max}$ for mixed micelles $C_i+C_{10}$ ($i$ = 14, 16, 18) as a function of the molar fraction $y_2$ of the shorter-chain surfactant ($C_{10}$) in the micelles: (a) Radius of a spherical micelle ($p = 1/3$); (b) radius of the endcaps of a spherocylindrical micelle at $p = 3/8$; (c) radius of a cylindrical micelle ($p = 1/2$), and (d) half-thickness of a lamellar micelle ($p = 1$).

For this goal, we first notice that $R$ must satisfy the following two inequalities:

$$R \leq l_1 \quad \text{and} \quad b \leq \frac{l_2}{R} \qquad (5.2)$$

The first inequality means that the micelle radius $R$ cannot exceed the length of the extended longer chains, $l_1$. The second inequality states that the dimensionless thickness of the outer region, $b$, cannot exceed the dimensionless length of the extended shorter chains, $l_2$; see Fig. 5. These two inequalities should be simultaneously satisfied. Hence, the maximal micelle radius, $R_{max}$, is determined by the relation

$$R_{max} \equiv \min(l_1, l_2/b) \qquad (5.3)$$

Because $b$ depends on the micelle composition and on the packing parameter $p$ (see Fig. 6), the same is true also for $R_{max}$.

As an illustration, Fig. 8 shows plots of $R_{max}$ vs. $y_2$ calculated from Eq. (5.3) for mixed micelles with chains $C_{14} + C_{10}$; $C_{16} + C_{10}$, and $C_{18} + C_{10}$, where the subscript $n$ in $C_n$ denotes the number of carbon atoms in the respective alkane chain. As usual, $y_2$ is the molar fraction of the surfactant of shorter chain (in this case $C_{10}$) in the micelles. Because of the relation $y_2 = 1 - y_1$, Fig. 8 gives the dependence of $R_{max}$ also on the mole fraction of the longer chains, $y_1$. The used chainlength values are those in Table 1.

As seen in Fig. 8, the micelle radius (half-thickness) varies between $l_1$ for the lower values of $y_2$ to $l_2$ for $y_2 = 1$. The kink in each curve corresponds to $l_2/b = l_1$, i.e. to $b = l_2/l_1$; see Eq. (5.3). For the same $l_2$, the kink is located at lower $y_2$ values for greater chainlengths $l_1$. Fig. 8 indicates that the effect of the shortest chains on $R_{max}$ is the strongest (takes place in the widest $y_2$ range) for lamellar micelles, whereas this effect is the weakest for spherical micelles.

It should be also noted that in the region with $R_{max} = l_2/b$, the curves in Fig. 8 are numerically close to the predictions of the semiempirical formula proposed in Ref [39].

*5.3. Free energy vs. micelle radius and composition*

Fig. 9 shows plots of the chain-conformational free energy per molecule, $f_{conf}$, vs. micelle radius, $R$, calculated using Eq. (4.18) along with Eqs. (4.16b) and (4.19). The calculations have been carried out for mixed micelles of components 1 and 2 with $C_{14}$ and $C_{10}$ chains, respectively, at three different molar fractions of the longer-chained surfactant, viz. $y_1 = 0.25$, 0.50 and 0.75.

The calculated curves correspond to three different values of the packing parameter: $p = 1/3$ (sphere); $p = 3/8$ (endcaps), and $p = 1/2$ (cylinder). Note that for $p = 1/3$ and $1/2$, $\beta_{conf}$ can be calculated from simpler analytical formulas, Eqs. (4.22) and (4.23), where the relation between the molar and volume fractions, $y_1$ and $\eta_1$, is given by Eq. (3.11).

Fig. 9 shows that for a given $R$, $f_{conf}$ increases in the order (sphere) < (endcap) < (cylinder). The molar fractions $y_1 = 0.25, 0.50, 0.75$ are *equidistant*. However, the results for the free energy are not equidistant, viz. $f_{conf}$ (0.75) ≈ $f_{conf}$ (0.50) < $f_{conf}$ (0.25). This pronouncedly nonlinear variation of $f_{conf}$ as a function of $\eta_1$ is related to the non-monotonic $\beta_{conf}$ ($\eta_1$) dependence in Fig. 7. This result is in agreement with experimental data for the dependence of the aggregation number of mixed nonionic wormlike micelles on their composition, as demonstrated in Ref. [49].

In Appendix B, it is demonstrated that the empirical formula for $f_{conf}$ proposed in Refs. [46,47] predicts systematically greater values of $f_{conf}$ in comparison with our theory.

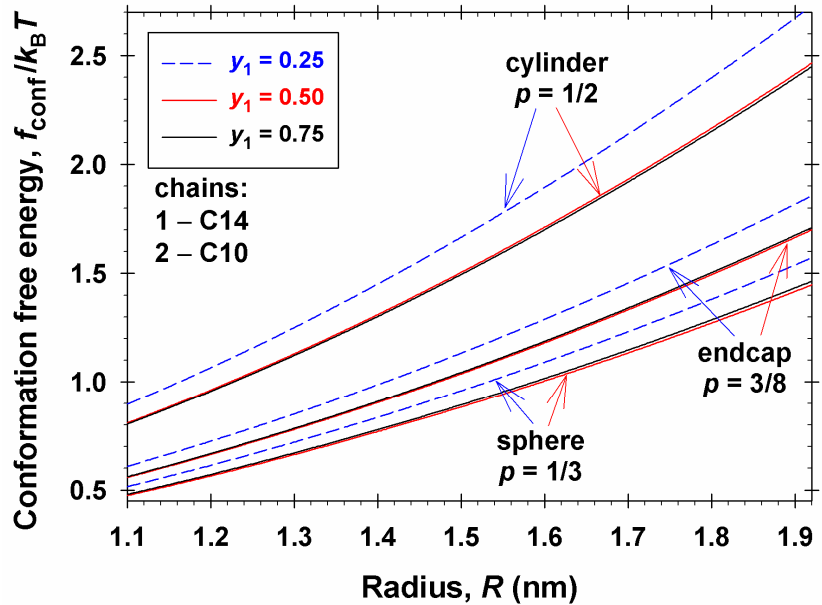

**Fig. 9**. Plots of the chain-conformational free energy per molecule, $f_{conf}$, vs. micelle radius, $R$, for mixed micelles with $C_{14}$ and $C_{10}$ chains at three different molar fractions of the longer-chained surfactant, $y_1 = 0.25, 0.50$ and 0.75, and for three different values of the packing parameter, $p = 1/3$ (sphere); $p = 3/8$ (endcap), and $p = 1/2$ (cylinder).

## 6. Conclusions

In this article, analytical expression for the chain-conformation free energy, $f_{conf}$, of *two-component* mixed micelles (spherical, wormlike and lamellar) is derived. The developed molecular-thermodynamic theory is a non-trivial generalization of the Semenov mean-field approach for single-component micelles [44], which was found to provide excellent quantitative description of the growth of wormlike micelles from nonionic surfactants [28].

By analytical minimization of the free-energy functional, we derived explicit expressions (i) for the chain-extension distribution functions, $\varepsilon_1(x,y)$ and $\varepsilon_2(x,y)$, and (ii) for the chain-end distribution functions, $g_1(y)$ and $g_2(y)$, in the case of two components of different chainlengths. The results indicate that the hydrocarbon core of a two-component micelle can be divided in two regions, *outer* and *inner*, where the ends of the shorter and longer chains are located (Fig. 5). The position, $b$, of the boundary between the two regions is determined as a function of the micelle composition and shape (Fig. 6). The obtained expression for the chain-conformation free energy, Eq. (4.18), indicates that the mixing of chains with different lengths is always *nonideal* and *synergistic* (Fig. 7), i.e. it leads to decrease of the micellar free energy, and therefrom – to enhancement of the micellization and micelle growth.

The chain-conformation free energy is an important component of the total micelle free energy. In the case of nonionic micelles, the other components are related to the interfacial-tension and headgroup-steric-repulsion effects [28]. In the second part of this study [49], we demonstrate that the combined theory, which includes the derived expression for $f_{conf}$, is in perfect quantitative agreement with experimental data for the aggregation number of mixed micelles from polyoxyethylene alkyl ethers without using any adjustable parameters. In particular, the combined theory predicts the experimentally observed nonlinear dependence of the micelle growth parameter on micelle composition, which is related to the nonlinear dependence of the chain-conformation free energy on composition (Fig. 9). In this respect, the molecular-thermodynamic theory developed in the present article represents a significant improvement over the semi-empirical expression for the chain-conformation free energy of mixed micelles proposed in Refs. [46,47], which is unable to describe the aforementioned nonlinear dependence.

The universality of the developed theory of the chain-conformation free energy of mixed micelles is related to the fact that it is applicable (i) to surfactants with different headgroups (nonionic, anionic, cationic and zwitterionic) and (ii) to micelles of different shapes (spherical, wormlike and lamellar). The developed methodology for two-component micelles could be further extended to micelles with three and more components. The derived theoretical expression for $f_{conf}$ could be incorporated in a quantitative molecular-thermodynamic theory of the growth of mixed wormlike micelles in formulations with various practical applications.

**Acknowledgements**

The authors gratefully acknowledge the support from Unilever R&D and from the Operational Programme ''Science and Education for Smart Growth", Bulgaria, project No. BG05M2OP001-1.001-0008.

**Appendix A and Appendix B**

Appendices A and B are attached as Supplementary Material to this article.

**References**

[1] S. Yoshimura, S. Shirai, Y. Einaga, Light-scattering characterization of the wormlike micelles of hexaoxyethylene dodecyl $C_{12}E_6$ and hexaoxyethylene tetradecyl $C_{14}E_6$ ethers in dilute aqueous solution, J. Phys. Chem. B 108 (2004) 15477–15487.

[2] S. Shirai, Y. Einaga, Wormlike micelles of polyoxyethylene dodecyl $C_{12}E_j$ and heptaoxyethylene alkyl $C_iE_7$ ethers. Hydrophobic and hydrophilic chain length dependence of the micellar characteristics, Polym. J. 37 (2005) 913–924.

[3] D. Danino, L. Abezgauz, I. Portnaya, N. Dan, From discs to ribbons networks: the second critical micelle concentration in nonionic sterol solutions, J. Phys. Chem. Lett. (2016) 1434–1439.

[4] Y. Yamashita, K. Savchuk, R. Beck, H. Hoffmann, Vesicle phases from alkyldimethylaminoxides and strong acids. The Hofmeister series of anions and the $L_1/L_\alpha$-phase transition, Curr. Opin. Colloid Interface Sci. 9 (2004) 173–177.

[5] Y. Yamashita, H. Maeda, H. Hoffmann, Counterion specificity in the phase behavior of tetradecyldimethylamine oxides at different degrees of protonation, J. Colloid Interface Sci. 299 (2006) 388–395.

[6] T. Imae, R. Kamiya, S. Ikeda, Formation of spherical and rodlike micelles of cetyltrimethylammonium bromide in aqueous NaBr solutions, J. Colloid Interface Sci. 108 (1985) 215–225.

[7] P.J. Missel, N.A. Mazer, M.C. Carey, G.B. Benedek, Influence of alkali-metal counterion identity on the sphere-to-rod transition in alkyl sulfate micelles, J. Phys. Chem. 93 (1989) 8354–8366.

[8] R. Alargova, J. Petkov, D. Petsev, I.B. Ivanov, G. Broze, A. Mehreteab. Light scattering study of sodium dodecyl polyoxyethylene-2-sulfate micelles in the presence of multivalent counterions, Langmuir 11 (1995) 1530–1536.

[9] M. Pleines, W. Kunz, T. Zemb, D. Banczédi, W. Fieber, Molecular factors governing the viscosity peak of giant micelles in the presence of salt and fragrances, J. Colloid Interface Sci. 537 (2019) 682–693.

[10] S. Hofmann, H. Hoffmann, Shear-induced micellar structures in ternary surfactant mixtures: The influence of the structure of the micellar interface, J. Phys. Chem. B 102 (1998) 5614–5624.

[11] N. Dan, K. Shimoni, V. Pata, D. Danino, Effect of mixing on the morphology of cylindrical micelles, Langmuir 22 (2006) 9860–9865.

[12] K. Imanishi, Y. Einaga. Wormlike micelles of polyoxyethylene alkyl ether mixtures $C_{10}E_5+C_{14}E_5$ and $C_{14}E_5+C_{14}E_7$: Hydrophobic and hydrophilic chain length dependence of the micellar characteristics, J. Phys. Chem. B 111 (2007) 62–73.

[13] F. Nettesheim, E.W. Kaler, Phase behavior of systems with wormlike micelles. In R. Zana, E. Kaler, eds. Giant Micelles. Properties and Applications, Taylor and Francis, New York, 2007, 41–79.

[14] H. Rehage, H. Hoffmann, Rheological properties of viscoelastic surfactant systems, J. Phys. Chem. 92 (1988) 4712–4719.

[15] H. Hoffmann, A. Rauscher, M. Gradzielski, S.F. Schulz, Influence of ionic surfactants on the viscoelastic properties of zwitterionic surfactant solutions, Langmuir 8 (1992) 2140–2146.

[16] S.R. Raghavan, G. Fritz, E.W. Kaler, Wormlike micelles formed by synergistic self-assembly in mixtures of anionic and cationic surfactants, Langmuir 18 (2002) 3797–3803.

[17] N.C. Christov, N.D. Denkov, P.A. Kralchevsky, K.P. Ananthapadmanabhan, A. Lips, Synergistic sphere-to-rod micelle transition in mixed solutions of sodium dodecyl sulfate and cocoamidopropyl betaine, Langmuir 20 (2004) 565–571.

[18] G. Colafemmina, R. Recchia, A.S. Ferrante, S. Amin, G. Palazzo, Lauric acid-induced formation of a lyotropic nematic phase of disk-shaped micelles, J. Phys. Chem. B 114 (2010) 7250–7260.

[19] M. Kamada, S. Shimizu, K. Aramaki, Manipulation of the viscosity behavior of wormlike micellar gels by changing the molecular structure of added perfumes, Colloids Surf. A 458 (2014) 110–116.

[20] L. Ziserman, L. Abezgauz, O. Ramon, S.R. Raghavan, D. Danino, Origins of the viscosity peak in wormlike micellar solutions. 1. Mixed catanionic surfactants. A cryo-transmission electron microscopy study, Langmuir 25 (2009) 10483–10489.

[21] S.E. Anachkov, P.A. Kralchevsky, K.D. Danov, G.S. Georgieva, K.P. Ananthapadmanabhan, Disclike vs. cylindrical micelles: Generalized model of micelle growth and data interpretation, J. Colloid Interface Sci. 416 (2014) 258–273.

[22] S. Chavda, D. Danino, V.K. Aswal, K. Singh, D.G. Marangoni, P. Bahadur, Microstructure and transitions in mixed micelles of cetyltrimethylammonium tosylate and bile salts, Colloids Surf. A 513 (2017) 223–233.

[23] J. Yang, Viscoelastic wormlike micelles and their applications, Curr. Opin. Colloid Interface Sci. 7 (2002) 276–281.

[24] S. Ezrahi, E. Tuval, A. Aserin, N. Garti, Daily applications of systems with wormlike micelles. In Giant Micelles. Properties and Applications. Zana R, Kaler EW, eds., CRC Press, 2007; 515–44.

[25] C.A. Dreiss, Wormlike micelles: where do we stand? Recent developments, linear rheology and scattering techniques, Soft Matter 3 (2007) 956–970.

[26] X. Tang, W. Zou, P.H. Koenig, S.D. McConaughy, M.R. Weaver, D.M. Eike, M. Schmidt, R.G. Larson, Multiscale modeling of the effects of salt and perfume raw materials on the rheological properties of commercial threadlike micellar solutions, J. Phys. Chem. B 121 (2017) 2468–2485.

[27] P.F. Sullivan, M.K.R. Panda, V. Laffite, Applications of wormlike micelles in oilfield industry. In Wormlike Micelles. Advances in Systems, Characterization and Applications. C.A. Dreiss, Y. Feng, eds., RSC, 2017; 330–352.

[28] K.D. Danov, P.A. Kralchevsky, S.D. Stoyanov, J.L Cook, I.P. Stott, E.G. Pelan, Growth of wormlike micelles in nonionic surfactant solutions: Quantitative theory vs. experiment, Adv. Colloid Interface Sci. 256 (2018) 1–22.

[29] Z. Mitrinova, S. Tcholakova, J. Popova, N. Denkov, B. Dasgupta, K.P. Ananthapadmanabhan, Efficient control of the rheological and surface properties of surfactant solutions containing C8−C18 fatty acids as cosurfactants, Langmuir 29 (2013) 8255–8265.

[30] G.S. Georgieva, S.E. Anachkov, I. Lieberwith, K. Koynov, P.A. Kralchevsky, Synergistic growth of giant wormlike micelles in ternary mixed surfactant solutions: Effect of octanoic acid, Langmuir 32 (2016) 12885–12893.

[31] S.E. Anachkov, G.S. Georgieva, L. Abezgauz, D. Danino, P.A. Kralchevsky, Viscosity peak due to shape transition from wormlike to disklike micelles: effect of dodecanoic acid, Langmuir 34 (2018) 4897–907.

[32] K.A. Dill, P.J. Flory, Interphases of chain molecules: Monolayers and lipid bilayer membranes, Proc. Natl. Acad. Sci. USA 77 (1980) 3115–3119.

[33] K.A. Dill, P.J. Flory, Molecular organization in micelles and vesicles, Proc. Natl. Acad. Sci. USA 78 (1981) 676–680.

[34] K.A. Dill, D.E. Koppel, R.S. Cantor, J.D. Dill, D. Bendedouch, S.-H. Chen, Molecular conformation in surfactant micelles, Nature 309 (1984) 42–45.

[35] A. Ben-Shaul, W.M. Gelbart, Theory of chain packing in amphiphilic aggregates, Ann. Rev. Phys. Chem. 36 (1985) 179–211.

[36] A. Ben-Shaul, I. Szleifer, W.M. Gelbart, Chain organization and thermodynamics in micelles and bilayers. I. Theory, J. Chem. Phys. 83 (1985) 3597–3611.

[37] I. Szleifer, A. Ben-Shaul, W.M. Gelbart, Chain organization and thermodynamics in micelles and bilayers. II. Model calculations, J. Chem. Phys. 83 (1985) 3612–3620.

[38] A. Ben-Shaul, D.H. Rorman, G.V. Hartland, W.M. Gelbart, Size distribution of mixed micelles: Rodlike surfactant-alcohol aggregates, J. Phys. Chem. 90 (1986) 5277–5286.

[39] I. Szleifer, A. Ben-Shaul, W.M. Gelbart, Statistical thermodynamics of molecular organization in mixed micelles and bilayers, J. Chem. Phys. 86 (1987) 7094–7109.

[40] S. May, A. Ben-Shaul, Molecular packing in cylindrical micelles. In R. Zana, E. Kaler, eds. Giant Micelles. Properties and Applications, Taylor and Francis, New York, 2007, 41–79.

[41] S. Puvvada, D. Blankschtein, Molecular-thermodynamic approach to predict micellization, phase behavior and phase separation of micellar solutions. I. Application to nonionic surfactants, J. Chem. Phys. 92 (1990) 3710–3724.

[42] S. Puvvada, D. Blankschtein, Theoretical and experimental investigation of micellar properties of aqueous solutions containing binary mixtures of nonionic surfactants, J. Phys. Chem. 96 (1992) 5579–5592.

[43] H.G. Thomas, A. Lomakin, D. Blankschtein, G.B. Benedek, Growth of mixed nonionic micelles, Langmuir 13 (1997) 209–218.

[44] A.N. Semenov, Contribution to the theory of microphase layering in block-copolymer melts, Sov. Phys. JETP 61 (1985) 733–742.

[45] R. Nagarajan, E. Ruckenstein, Theory of surfactant self-assembly: A predictive molecular thermodynamic approach, Langmuir 7 (1991) 2934–2969.

[46] R. Nagarajan, Micellization of binary surfactant mixtures. Theory. In P.M. Holland, D.N. Rubingh, eds., Mixed Surfactant Systems, ACS Symposium Series, vol. 501, Washington, DC, 1992, 54–95.

[47] R. Nagarajan, Molecular thermodynamics of giant micelles, In R. Zana, E. Kaler, eds. Giant Micelles. Properties and Applications, Taylor and Francis, New York, 2007, 1–40.

[48] M.S. Kshevetskiy, A.K. Shchekin, The aggregation work and shape of molecular aggregates upon the transition from spherical to globular and cylindrical micelles, Colloid J. 67 (2005) 324–336.

[49] K.D. Danov, P.A. Kralchevsky, S.D. Stoyanov, J.L Cook, I.P. Stott, Analytical modeling of micelle growth. 2. Molecular thermodynamics of mixed aggregates and scission energy in wormlike micelles, J. Colloid Interface Sci. (2019) – in press.

[50] J.N. Israelachvili, Intermolecular and Surface Forces, 3$^{rd}$ ed. Amsterdam, Academic Press, 2011.

[51] P.J. Flory, Principles of Polymer Chemistry. Ithaca, NY, Cornell University Press, 1962.

[52] C. Tanford, The Hydrophobic Effect. The Formation of Micelles and Biological Membranes, 2$^{nd}$ ed., New York, Wiley, 1980.

# Supplementary material

for the article

# Analytical modeling of micelle growth. 1. Chain-conformation free energy of binary mixed spherical, wormlike and lamellar micelles

Authors: Krassimir D. Danov, Peter A. Kralchevsky, Simeon D. Stoyanov, Joanne L. Cook, and Ian P. Stott

The cited equations and references have the same numbers as in the main text of the article.

## Appendix A. Analytical solution of the variational problem

The integration of Eq. (3.8) yields

$$\int_0^1 \frac{s_k(y)}{p}(1-y)^{\frac{1-p}{p}}\,\mathrm{d}y = \eta_k \quad (k=1,\,2) \tag{A.1}$$

where we have used Eqs. (3.4) and (3.5), as well as a change of the order of integration in accordance with the formula

$$\int_0^a \mathrm{d}x \int_0^x \mathrm{d}y\, f(x,y) = \int_0^a \mathrm{d}y \int_y^a \mathrm{d}x\, f(x,y) \tag{A.2}$$

(see Fig. A.1) where $f(x,y)$ is an arbitrary integrand. To derive Eq. (A.1), we have set $a = 1$ and $x = x_k$. One could verify that Eq. (A.1) is consistent with the identities $\eta_1 + \eta_2 = 1$ and $s_1(y) + s_2(y) = 1$.

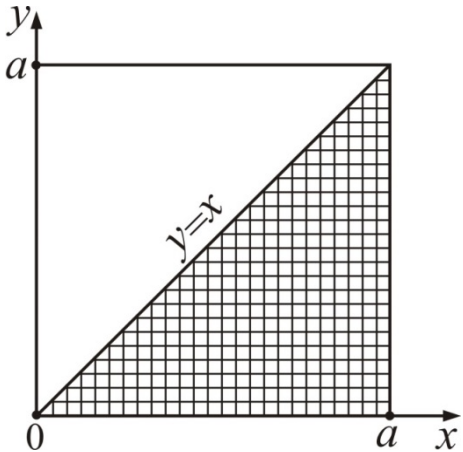

**Fig. A.1**. The double integral over the hatched triangle can be expressed in two equivalent ways, see Eq. (A.2).

In the minimization procedure, it is assumed that the functions $s_1(y)$ and $s_2(y)$ satisfy Eq. (A.1) by definition. In such a case, Eq. (3.4) becomes a corollary from Eqs. (3.5) and (3.8), and does not represent an independent constraint.

Here, our goal is to minimize the Lagrange functional $\Phi$ defined by Eq. (3.13) at fixed functions $s_1(y)$ and $s_2(y)$. At the minimum, the first variation of $\Phi$ has to be equal to zero, and correspondingly, the coefficients before the independent variations of the functions $g_k$, $\varepsilon_k$, $\lambda_k$ and $\gamma_k$, $k = 1, 2$, should be set equal to zero. This leads to a system of equations for determining these functions.

The coefficients before the variations of the Lagrange multipliers $\lambda_k$ and $\gamma_k$ give the constraints expressed by Eqs. (3.5) and (3.8). The coefficient before the variation of $g_k$ yields:

$$\int_0^{x_k} [\varepsilon_k(x_k, y) - \frac{\gamma_k(y)}{\varepsilon_k(x_k, y)}] dy = 0 \quad (k = 1, 2) \tag{A.3}$$

To derive Eq. (A.3), we have used a change of the order of integration in accordance with Eq. (A.2).

The coefficient before the variation of $\varepsilon_k$ yields:

$$-g_k(x_k)\varepsilon_k^2(x_k, y) - \lambda_k(x_k) - \gamma_k(y)g_k(x_k) = 0 \quad (k = 1, 2) \tag{A.4}$$

Following Semenov [44], we assume that the end segments of all surfactant chains are not extended, that is $\varepsilon_k(x_k, x_k) = 0$. Then, setting $y = x_k$ in Eq. (A.4) we obtain:

$$-\lambda_k(x_k) = \gamma_k(x_k)g_k(x_k) \quad (k = 1, 2) \tag{A.5}$$

The substitution of Eq. (A.5) in Eq. (A.4) yields:

$$\varepsilon_k^2(x_k, y) = \gamma_k(x_k) - \gamma_k(y) \quad (k = 1, 2) \tag{A.6}$$

Furthermore, the substitution of Eq. (A.6) in the constraint expressed by Eq. (3.5) leads to an Abel type integral equation:

$$\int_0^{x_k} \frac{dy}{[\gamma_k(x_k) - \gamma_k(y)]^{1/2}} = 1 \quad \text{for} \quad 0 < x_k \leq 1 \quad (k = 1, 2) \tag{A.7}$$

The solution of the last integral equation reads:

$$\gamma_k(y) = \gamma_k(0) + (\frac{\pi}{2})^2 y^2 \quad (k = 1, 2) \tag{A.8}$$

The substitution of Eq. (A.8) in Eq. (A.6) gives an explicit expression for the dimensionless local extension function:

$$\varepsilon_k(x_k, y) = \frac{\pi}{2}(x_k^2 - y^2)^{1/2} \quad (k = 1, 2) \tag{A.9}$$

In addition, the substitution of Eqs. (A.8) and (A.9) into Eq. (A.3) yields $\gamma_k(0) = 0$.

Furthermore, the substitution of Eq. (A.9) into the constraint expressed by Eq. (3.8) leads to an integral equation for determining the function $g_k(x)$:

$$\frac{\pi}{2p} s_k(y)(1-y)^{\frac{1-p}{p}} = \int_y^1 \frac{g_k(x)}{(x^2-y^2)^{1/2}} dx \quad \text{for} \ \ 0 \leq y \leq 1 \ \ (k=1,2) \tag{A.10}$$

Eq. (A.10) is again an Abel type integral equation. Its solution is given by Eq. (3.15); see: A. Chakrabarti, Solution of the generalized Abel integral equation, *J. Integral Equ. Appl.* 20 (2008) 1–11.

## Appendix B. Additional theoretical derivations

*B.1 Investigation for possible additional constraints on the variations of $g_2(x)$*

Eq. (A.10) for $k = 2$ can be represented in the form:

$$s_2(y) = \frac{2p}{\pi}(1-y)^{\frac{p-1}{p}} \int_y^1 \frac{g_2(x)}{(x^2-y^2)^{1/2}} dx \quad \text{for} \ \ 0 \leq y \leq 1 \tag{B.1}$$

By definition, $0 \leq s_2(y) \leq 1$. Hence, the integral in Eq. (B.1) should satisfy the same inequality:

$$0 \leq \frac{2p}{\pi}(1-y)^{\frac{p-1}{p}} \int_y^1 \frac{g_2(x)}{(x^2-y^2)^{1/2}} dx \leq 1 \quad \text{for} \ \ 0 \leq y \leq 1 \tag{B.2}$$

In view of Eq. (4.3) in the main text, we have $0 \leq g_2(x) \leq g(x)$, and consequently

$$0 \leq \frac{g_2(x)}{(x^2-y^2)^{1/2}} \leq \frac{g(x)}{(x^2-y^2)^{1/2}} \quad \text{for} \ \ 0 < x \leq 1 \ \text{and} \ y < x \tag{B.3}$$

By integration of Eq. (B.3) with respect to $x$, we derive:

$$0 \leq \frac{2p}{\pi}(1-y)^{\frac{p-1}{p}} \int_y^1 \frac{g_2(x)}{(x^2-y^2)^{1/2}} dx \leq \frac{2p}{\pi}(1-y)^{\frac{p-1}{p}} \int_y^1 \frac{g(x)}{(x^2-y^2)^{1/2}} dx \quad \text{for} \ \ 0 \leq y \leq 1 \tag{B.4}$$

Having in mind that $s_1 + s_2 = 1$ and $g_1 + g_2 = g$, the summation of the two expressions in Eq. (A.10) yields:

$$\int_y^1 \frac{g(x)}{(x^2-y^2)^{1/2}} dx = \frac{\pi}{2p}(1-y)^{\frac{1-p}{p}} \quad \text{for} \ \ 0 \leq y \leq 1 \tag{B.5}$$

The substitution of Eq. (B.5) into Eq. (B.4) leads to Eq. (B.2). Hence, Eq. (B.2) turns out to be a corollary of Eq. (4.3), so that it does not impose any additional constraint on the variations of $g_2(x)$.

## B.2 Chain-conformation free energy, $f_{conf}$, and interaction coefficient, $\beta_{conf}$

To transform the last integral in Eq. (4.17), we substitute $g(x)$ from Eq. (3.16) and integrate by parts:

$$\int_0^b g(x)x^2\,dx = -\int_0^b dx\, x^2 \frac{d}{dx}[\int_x^1 dz\, \frac{z(1-z)^{\frac{1-p}{p}}}{p(z^2-x^2)^{1/2}}]$$

$$= -b^2\int_b^1 dz\, \frac{z(1-z)^{\frac{1-p}{p}}}{p(z^2-b^2)^{1/2}} + 2\int_0^b dx\, x[\int_x^1 dz\, \frac{z(1-z)^{\frac{1-p}{p}}}{p(z^2-x^2)^{1/2}}]$$

$$= -b^2\eta_1 + 2\int_0^b dx\, x[\int_x^b dz\, \frac{z(1-z)^{\frac{1-p}{p}}}{p(z^2-x^2)^{1/2}}] + 2\int_0^b dx\, x[\int_b^1 dz\, \frac{z(1-z)^{\frac{1-p}{p}}}{p(z^2-x^2)^{1/2}}] \quad (B.6)$$

see Eq. (4.12). Further, using Eq. (A.2) we change the order of integration in Eq. (B.6) and take the integrals with respect to $x$:

$$\int_0^b g(x)x^2\,dx = -b^2\eta_1 + \frac{2}{p}\int_0^b dz\, z(1-z)^{\frac{1-p}{p}} \int_0^z dx\, \frac{x}{(z^2-x^2)^{1/2}} + \frac{2}{p}\int_b^1 dz\, z(1-z)^{\frac{1-p}{p}} \int_0^b dx\, \frac{x}{(z^2-x^2)^{1/2}}$$

$$= -b^2\eta_1 + \frac{2}{p}\int_0^b z^2(1-z)^{\frac{1-p}{p}}\,dz + \frac{2}{p}\int_b^1 z(1-z)^{\frac{1-p}{p}}[z-(z^2-b^2)^{1/2}]\,dz \quad (B.7)$$

Furthermore, we substitute Eq. (B.7) into Eq. (4.17) and take into account the definition of $c_{conf}(p)$ given by Eq. (4.16b):

$$\frac{f_{conf}}{k_B T} = \frac{3\pi^2 R^2 \bar{l}}{16 l_{sg}} \left\{ \frac{c_{conf}(p)}{l_2^2} - (\frac{1}{l_2^2} - \frac{1}{l_1^2})\left[ b^2\eta_1 + \frac{2}{p}\int_b^1 z(1-z)^{\frac{1-p}{p}}(z^2-b^2)^{1/2}\,dz \right] \right\} \quad (B.8)$$

Next, in Eq. (B.8) we separate the terms corresponding to *ideal* and *nonideal* mixing, and obtain Eqs. (4.18) and (4.19); see also Eq. (4.21).

(a) *In the case of lamella*, we have $p = 1$ and $c_{conf}(1) = 2/3$; see Eq. (4.16b). The integral in Eq. (B.8) reduces to

$$\int_b^1 (z^2-b^2)^{1/2} z\,dz = \frac{1}{3}(1-b^2)^{3/2} \quad (B.9)$$

The substitution of Eqs. (B.9) and (4.14) in the expression for $\beta_{conf}$, Eq. (4.19), after some transformations leads to

$$\beta_{conf}(1,\eta_1) = \frac{\eta_1}{2}(1-\eta_1^2) \qquad (B.10)$$

which is Eq.(4.24) in the main text; $\beta_{conf}$ defined by Eq. (B.10) is positive and has a maximum with coordinates (see Fig. 7):

$$\eta_1 = \sqrt{3}/3 = 0.57735, \quad \beta_{conf} = \sqrt{3}/9 \approx 0.19245 \qquad (B.11)$$

(b) *In the case of cylinder*, we have $p = 1/2$ and the integral in Eq. (4.12) can be taken analytically, so that it acquires the form:

$$(1-b^2)^{1/2} - b^2 \ln[\frac{1}{b} + (\frac{1}{b^2}-1)^{1/2}] = \eta_1 \qquad (B.12)$$

Likewise, taking the integral in Eq. (B.8) we obtain:

$$\frac{2}{p}\int_b^1 z(1-z)^{\frac{1-p}{p}}(z^2-b^2)^{1/2}\,dz = \frac{b^4}{2}\ln[\frac{1}{b}+(\frac{1}{b^2}-1)^{1/2}] + \frac{2-5b^2}{6}(1-b^2)^{1/2} \text{ for } p=1/2 \qquad (B.13)$$

Eliminating the logarithmic terms between Eqs. (B.12) and (B.13), we get:

$$\frac{2}{p}\int_b^1 z(1-z)^{\frac{1-p}{p}}(z^2-b^2)^{1/2}\,dz + b^2\eta_1 = \frac{1}{3}(1-b^2)^{3/2} + \frac{b^2}{2}\eta_1 \text{ for } p=1/2 \qquad (B.14)$$

In view of Eq. (4.12), we have $c_{conf}(1/2) = 1/3$, so that

$$\frac{1}{c_{conf}(p)}[\frac{2}{p}\int_b^1 z(1-z)^{\frac{1-p}{p}}(z^2-b^2)^{1/2}\,dz + b^2\eta_1] = (1-b^2)^{3/2} + \frac{3}{2}b^2\eta_1 \text{ for } p=1/2 \qquad (B.15)$$

Substituting Eq. (B.15) in Eq. (4.19), we obtain:

$$\beta_{conf}(\frac{1}{2},\eta_1) = (1-b^2)^{3/2} + \frac{3}{2}b^2\eta_1 - \eta_1 \qquad (B.16)$$

which is Eq.(4.23) in the main text; $\beta_{conf}$ defined by Eq. (B.16) is positive and has a maximum with coordinates (see Fig. 7):

$$\eta_1 = (\frac{2}{3})^{1/2} - \frac{1}{3}\ln(\sqrt{2}+\sqrt{3}) \approx 0.434425, \quad \beta_{conf} = \frac{\sqrt{6}}{18} + \frac{1}{6}\ln(\sqrt{2}+\sqrt{3}) \approx 0.327119 \qquad (B.17)$$

(c) *In the case of sphere*, we have $p = 1/3$ and the integral in Eq. (4.12) can be taken analytically, so that it acquires the form:

$$(1-b^2)^{1/2}(1+2b^2) - 3b^2 \ln[\frac{1}{b}+(\frac{1}{b^2}-1)^{1/2}] = \eta_1 \qquad (B.18)$$

Likewise, taking the integral in Eq. (B.8) we obtain:

$$\frac{2}{p}\int_b^1 z(1-z)^{\frac{1-p}{p}}(z^2-b^2)^{1/2}\,dz = \frac{3b^4}{2}\ln[\frac{1}{b}+(\frac{1}{b^2}-1)^{1/2}]+\frac{2-9b^2-8b^4}{10}(1-b^2)^{1/2} \quad \text{for } p=1/3 \quad \text{(B.19)}$$

Eliminating the logarithmic terms between Eqs. (B.18) and (B.19), we get:

$$\frac{2}{p}\int_b^1 z(1-z)^{\frac{1-p}{p}}(z^2-b^2)^{1/2}\,dz + b^2\eta_1 = \frac{1}{5}(1-b^2)^{5/2}+\frac{\eta_1}{2}b^2 \quad \text{for } p=1/3 \quad \text{(B.20)}$$

In view of Eq. (4.12), we have $c_{\text{conf}}(1/3) = 1/5$, so that

$$\frac{1}{c_{\text{conf}}(p)}[\frac{2}{p}\int_b^1 z(1-z)^{\frac{1-p}{p}}(z^2-b^2)^{1/2}\,dz + b^2\eta_1] = (1-b^2)^{5/2}+\frac{5}{2}b^2\eta_1 \quad \text{for } p=1/3 \quad \text{(B.21)}$$

Substituting Eq. (B.21) in Eq. (4.19), we obtain:

$$\beta_{\text{conf}}(\frac{1}{3},\eta_1) = (1-b^2)^{5/2}+\frac{5}{2}b^2\eta_1 - \eta_1 \quad \text{(B.22)}$$

which is Eq. (4.22) in the main text; $\beta_{\text{conf}}$ defined by Eq. (B.22) is positive and has a maximum with coordinates (see Fig. 7):

$$\eta_1 = \frac{14}{5^{3/2}}-\frac{3}{5}\ln(2+\sqrt{5}) \approx 0.386017, \quad \beta_{\text{conf}} = \frac{3}{10}\ln(2+\sqrt{5})-\frac{3}{5^{5/2}} \approx 0.379425 \quad \text{(B.23)}$$

*B.3 Comparison with the semiempirical expression for $f_{\text{cor}}$*

Based on expressions from the Semenov mean-field theory for single-component micelles [44], Nagarajan [46,47] proposed the following semiempirical expressions for the chain-conformation free energy (per molecule) of binary mixed micelles:

$$\frac{f_{\text{conf}}}{k_B T} = \frac{\lambda \pi^2 p}{80}\left(y_1\frac{R^2}{l_{sg}l_1}+y_2\frac{Q^2}{l_{sg}l_2}\right) \quad \text{(B.24)}$$

where, as usual, $y_1$ and $y_2$ are the molar fractions of the two surfactants in the micelle; $l_1$ and $l_2$ are their chainlengths; $l_{sg}$ is the length per segment; $R$ is the radius of the micelle hydrocarbon core; $p$ is the packing parameter, and

$$\lambda = \begin{cases} 9 & \text{for } 1/3 \le p \le 3/8; \\ 10 & \text{for } p=1/2; \end{cases} \quad Q = \begin{cases} R & \text{for } R<l_1,l_2; \\ l_2 & \text{for } l_2 \le R \le l_1. \end{cases} \quad \text{(B.25)}$$

In Fig. B.1, we compare the values of $f_{conf}$ calculated from our Eq. (4.18), with the predictions of the semiempirical Eq. (B.24). The calculation is carried out for a mixture, in which components 1 and 2 have $C_{14}$ and $C_{10}$ chains, respectively, at three molar fractions: $y_1 = 0.25$, 0.50 and 0.75. Fig. B.1 indicates that the semiempirical Eq. (B.24) predicts considerably greater values of $f_{conf}$ as compared to Eq. (4.18). Note that even a difference of the order of 0.1 $k_BT$ in $f_{conf}$ is essential for the micelle growth [28].

Moreover, Eq. (B.24) predicts an almost linear variation of $f_{conf}$ with $y_1$ (equidistant curves), whereas Eq. (4.18) predicts a pronouncedly nonlinear variation of $f_{conf}$ with $y_1$ (the curves for $y_1 = 0.50$ and 0.75 almost coincide), in agreement with the experimental data [49].

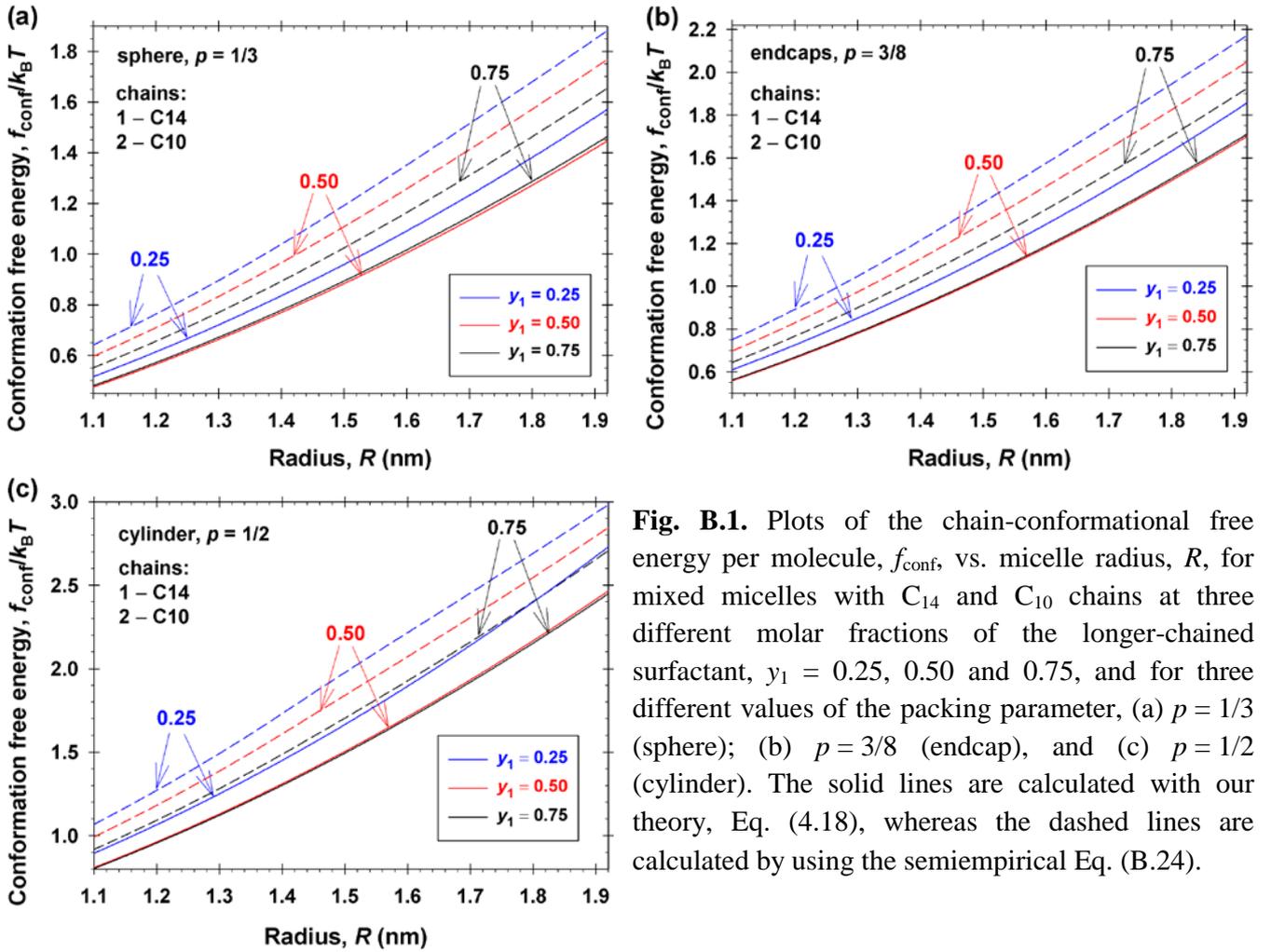

**Fig. B.1.** Plots of the chain-conformational free energy per molecule, $f_{conf}$, vs. micelle radius, $R$, for mixed micelles with $C_{14}$ and $C_{10}$ chains at three different molar fractions of the longer-chained surfactant, $y_1 = 0.25$, 0.50 and 0.75, and for three different values of the packing parameter, (a) $p = 1/3$ (sphere); (b) $p = 3/8$ (endcap), and (c) $p = 1/2$ (cylinder). The solid lines are calculated with our theory, Eq. (4.18), whereas the dashed lines are calculated by using the semiempirical Eq. (B.24).